\documentclass[a4paper,fleqn,usenatbib]{mnras}
\usepackage{amsmath}	
\usepackage{amssymb}	
\providecommand{\abs}[1]{\lvert#1\rvert}
\usepackage{newtxtext,newtxmath}

\usepackage[T1]{fontenc}
\usepackage{ae,aecompl}
\usepackage{hyperref}

\usepackage{graphicx}	
\newcommand{\vecR}{\mbox{\boldmath $R$} {}}
\newcommand{\vecd}{\mbox{\boldmath $d$} {}}
\newcommand{\vecv}{\mbox{\boldmath $v$} {}}
\newcommand{\vecV}{\mbox{\boldmath $V$} {}}
\newcommand{\vecL}{\mbox{\boldmath $L$} {}}

\newcommand{\vece}{\mbox{\boldmath $e$} {}}
\newcommand{\vecr}{\mbox{\boldmath $r$} {}}
\newcommand{\vecomega}{\mbox{\boldmath $\Omega$} {}}
\newcommand{\vecna}{\mbox{\boldmath $\nabla$}{}}

\title[Gaps by inclined planets ]{Gap formation by inclined massive planets in locally isothermal three-dimensional discs}

\author[R. O. Chametla et al.]{
Ra\'ul O. Chametla,$^{1}$\thanks{E-mail:rortegac0500@alumno.ipn.mx}
F. J. S\'anchez-Salcedo$^{2}$,
F. S. Masset$^{3}$,
and A. M. Hidalgo-G\'amez$^{1}$
\\
$^{1}$Escuela Superior de F\'{\i}sica y Matem\'aticas, Instituto Polit\'ecnico Nacional, U. P.
Adolfo L\'opez Mateos, Zacatenco, 07738 Mexico City, \\Mexico\\
$^{2}$Instituto de Astronom\'{\i}a, Universidad Nacional Aut\'onoma de M\'exico, Ciudad
Universitaria, Apt. Postal 70-264, C.P. 04510,
\\Mexico City, Mexico\\
$^{3}$Instituto de Ciencias F\'{\i}sicas, Universidad Nacional
Aut\'onoma de M\'exico, C.P. 62210 Cuernavaca,
Morelos,  Mexico
}

\date{Accepted XXX. Received YYY; in original form ZZZ}

\pubyear{2015}

\begin{document}
\label{firstpage}
\pagerange{\pageref{firstpage}--\pageref{lastpage}}
\maketitle

\begin{abstract}
  We study gap formation in gaseous protoplanetary discs by a Jupiter mass planet.
  The planet's orbit is circular and inclined relative to the midplane of the disc. We use 
  the impulse approximation to estimate the gravitational tidal torque between the
  planet and the disc, and infer the gap profile. For low-mass discs, we provide a criterion for gap 
  opening when the orbital inclination is $\leq 30^{\circ}$.  
  Using the FARGO3D code, we simulate the disc response
  to an inclined massive planet. The dependence of the depth and width of the
  gap obtained in the simulations on the inclination of the planet is
  broadly consistent with the scaling laws derived in the impulse
  approximation.  Although we mainly focus on planets kept on fixed orbits,  
  the formalism permits to infer the temporal evolution of the gap profile in cases where
  the inclination of the planet changes with time. This study may be useful to understand the 
  migration of massive planets on inclined orbit, because the strength
  of the interaction with the disc depends on whether a gap is opened
  or not.

\end{abstract}




\section{Introduction}
 \label{sec:intro}
Models of planetary formation that involve either core accretion or fragmentation of protoplanetary discs predict
that the orbit of the planets should lie in the disc \citep{Pollacketal1996, Mayeretal2002}. There is a large body of work on the 
tidal interaction between a planet and the protoplanetary
disc assuming that the planet is orbiting in the midplane of the disc \citep{Lin1993, Brydenetal1999, Varniereetal2004, Armitage2010, Kley2012, Baruteau2013}.
The density perturbations in the protoplanetary disc exert a tidal torque on the planet,
so it may migrate radially. Low mass planets (with masses below a few to a few tens of Earth masses) 
induce linear perturbations in the structure of the disc, whereas more massive
planets produce non-linear perturbations.
In the latter case,
the transfer of angular momentum from the planet to the disc may lead to the opening
of a gap in the disc. 
Interestingly, the existence of gaps in discs around very young stars (type HL Tau) has been recently
confirmed in submillimeter observations with the Atacama Large Millimeter-Submillimeter
Array (ALMA) \citep[e.g.,][]{Carrasco-Gonzalezetal2016, Hsi-WeiYenetal2016}. Whether these
gaps are created by massive planets or not is still under debate.

Until this day, about $3300$ extrasolar planets have been detected through either radial
velocity or transit measurements. Using the Rossiter-McLaughlin effect \citep{Fabrycky2009} it is possible
to calculate the tilt angle between the sky projection of the stellar spin axis and
the spin axis of the orbit of the planet. It was found that $40\%$ of the massive planets observed have a non-zero
tilt angle \citep{Triaudetal2010,Albrecht2012}. Different hypothesis have been suggested to explain
how planets can have misaligned orbits \citep[e.g.,][]{Xiang2013,Picogna2015}. 

The evolution of the orbital parameters of a planet on an inclined orbit due to its interaction with the
protoplanetary disc through tidal torques has been investigated by
several authors. For low-mass planets on orbits with eccentricity and
inclination smaller than the disc's aspect ratio,
\citet{TanakaWard2004} performed linear calculations and predicted a rapid
exponential decay of the inclination $i$ and eccentricity $e$ of the
planetary orbit \citep[see also][]{CresswellNelson2006}. 
For larger initial values of $e$ and $i$, the orbital evolution of a $20$ Earth-mass planet
was studied numerically by \citet{Cresswelletal2007}, who found that 
the time scales for eccentricity and inclination damping, albeit longer than given by the linear analysis of \citet{TanakaWard2004}, are still shorter than the migration time scale. 

The orbital evolution of massive planets is more complex.
\citet{MarzariNelson2009} studied the orbital evolution of a  Jupiter mass ($M_{J}$) planet with
an initial inclination of $20^{\circ}$ and initial eccentricities ranging from $0$ to $0.4$.
For an isothermal disc with a local surface density at the planetary orbit of $242$ g cm$^{-2}$,
they found that the inclination and eccentricity are rapidly damped on a timescale
of the order of $10^{3}$ years.
\citet{Xiang2013} considered the orbital evolution of planets between $1M_{J}$ and $6M_{J}$,
initialized with zero eccentriciy and a wide range of inclinations.
They showed that the inclination decay rate decreases drastically with
the initial inclination. For instance, for a Jupiter mass planet with an initial
inclination of $80^{\circ}$, the time required for inclination to decay by $10^{\circ}$
is of the order of $10^{6}$ years \citep[see also][]{Rein2012}. \citet{Bitsch2013} also investigated 
numerically the evolution of inclination and eccentricity for planets above $1M_{J}$ 
and provided empirical formulae for $di/dt$ and $de/dt$ by fitting the results
of their simulations. \citet{Lubow2015} and \citet{Miranda2015} investigated the tidal truncation
of misaligned discs in binary systems by computing the Lindblad torques.

Many of the previous studies focused mainly on determining whether inclined planets can, or cannot, realign with the protoplanetary disc within the lifetime of the disc.  The three-dimensional structure of the disc also changes because tidal torques by an inclined planet can open gaps in the disc, excite bending waves or warps, and can make them eccentric.  Here we are interested in the gap clearing by massive inclined planets.  Gap opening has been studied thoroughly in the coplanar case because the planet migration and the mass accretion rates are both sensitive to the existence of a gap.  Less studied is the gap opening by inclined planets. 
Simulations indicate that planets with low inclinations produce much wider and deeper gaps than
planets with large inclinations \citep{Xiang2013, Bitsch2013}. However, there is no physical description of the width and shape of these gaps.  As occurs in the coplanar case, \citet{Xiang2013} noticed that for small and
intermediate inclinations, the rate of inclination damping depends on gap formation; it decreases as soon as the gap is formed because the strength of the interaction with the disc depends on the local disc density.

Given the recent observations of gaps in circumstellar discs and given the importance of gaps to understand the orbital decay of planets and the gas accretion onto giant protoplanets, we study the gap formation by a Jupiter mass planet on an inclined orbit relative to the initial midplane of the disc, when the inclination is $30^{\circ}$ or lower.

This paper is organized as follows. In Section \ref{sec:cop_case}, we
review the basics of gap opening in the coplanar case.  In Section
\ref{sec:differential_ring}, a model based on the impulse approximation is
presented for calculating the torque between the disc and the planet.
Section \ref{sec:gapsmethods} describes the methods to derive the gap profile. In Section
\ref{sec:num_sim} we present our simulations, show the
three-dimensional (3D) structure of the disc and compare the resulting
gap profile to our analytical model.  Finally, our main
conclusions are given in Section \ref{sec:conclusions}.

\section{The coplanar case: Torques and gap formation criteria}
 \label{sec:cop_case}

Consider a thin disc with a smooth surface density $\Sigma (R)$ rotating with Keplerian angular frequency
$\Omega(R)$ around a star of mass $M_{S}$, and a planet of mass $M_{p}$ on circular
orbit with radius $R_{p}$. If the mass of the planet is sufficiently high,
the tidal torque on the disc can open a gap in the vicinity of the
planet's orbit. In the coplanar case, if the disc has a gap, there are four relevant scale lengths: 
the orbital radius $R_{p}$,
the thickness of the disc ($H$), the Hill radius $r_{H}\equiv (M_{p}/3M_{S})^{1/3}R_{p}$,
and the distance between the orbit of the planet and the edge of the gap $\Delta_{0}$.
From simple physical grounds, one expects the following ordering between these
scales:
\begin{equation}
\Delta_{0}\gtrsim H \hskip 0.2cm {\rm and} \hskip 0.2cm \Delta_{0}\gtrsim r_{H}
\label{eq:ordering}
\end{equation} 
\citep[e.g.,][]{Lin1993}.

The one-sided torque between the protoplanet 
and the disc when they are coplanar, denoted by $T_{g}$, has been derived
using different approaches \citep[see][for a review]{Lin1993}. 
Using the impulse approximation, 
\citet{LinPapaloizou1979} obtained
\begin{equation}
T_{g}=C_{T} q^2\Sigma R_p^4 \omega^2 \left(\frac{R_p}{\Delta_{0}}\right)^3,
\label{eq:Tg_impulse_approx_coplanar}
\end{equation}
where $q$ is the planet to star mass ratio ($q\equiv M_{p}/M_{S}$),
$\omega\equiv \Omega(R_{p})$ is the angular frequency of the planet, 
and $C_{T}=8/27$. 

 Alternatively, $T_{g}$ can be also calculated by adding the
contribution of the torques exerted on the disc at all the Lindblad resonances 
\citep[e.g.,][]{GoldreichTremaine1980,Ward1986, Lin1993}. In this formalism,  the Equation
(\ref{eq:Tg_impulse_approx_coplanar}) is recovered with $C_{T}=(32/243)[2K_{0}(2/3)+K_{1}(2/3)]^{2}
\simeq 0.84$ (where $K_{0}$ and $K_{1}$ are modified Bessel functions; see, e.g., equation (21) in
Lin \& Papaloizou 1993).

\citet{Papaloizou1984} obtained $T_{g}$ by computing the angular 
momentum transfered between fluid elements and the planet, using the WKB approximation
and taking into account the truncated disc structure.
They found that the gravitational torque is maximum when $\Delta_{0}\simeq H$,  
and that the maximum value is  
\begin{equation}
    T_{g} =  0.23 q^2\Sigma R_p^4 \omega^2 \left(\frac{R_p}{H}\right)^3,
	\label{eq:H_tot}
\end{equation}
where $\Sigma$ is the surface density outside to the gap
(in practice, it is usually taken as the unperturbed density at the planet radius, which will
be denoted by $\Sigma_{0}$). Note that the above equation is in agreement with 
Equation (\ref{eq:Tg_impulse_approx_coplanar}) with $C_{T}$ as derived
in the impulse approximation, provided that $H\simeq \Delta_{0}$.
In terms of the aspect ratio $h\equiv H/R_{p}$, we can write
$T_{g} =  0.23 q^2\Sigma_{0} R_p^4 \omega^2 h^{-3}$.

On the other hand, the angular momentum flux due to viscous stresses in a Keplerian disc with constant viscosity 
$\nu$ is given by 
\begin{equation}
   T_{\nu} =3\pi\Sigma\nu R^2\Omega,
	\label{eq:Tensor_visc}
\end{equation}
\citep[e.g.,][]{Lin1993}.
Equating $T_{g}$ and $T_{\nu}$, 
and assuming the ordering given in Equation (\ref{eq:ordering}), the viscous 
condition for the gap formation is given as
\begin{equation}
   q\gtrsim q_{\rm crit}\equiv \frac{40\nu}{\omega R^2_p}.
	\label{eq:criteria_LP93}
\end{equation}
\citet{Brydenetal1999} found through numerical simulations that a {\it clean, deep}
gap forms if $q>q_{\rm crit}$ \citep[see also][]{Lin1993}. For a typical disc with $h=0.05$,
simulations showed that even for $q=q_{\rm crit}$, the surface density at the bottom of
the gap is $\sim 0.2\Sigma_{0}$ \citep[e.g.,][]{hos07}.

\citet{Cridaetal2006}, based on a semi-analytic study, obtained a more general gap opening criterion by considering a pressure torque in addition to the viscosity and gravity torques. This criterion involves simultaneously the planet mass, viscosity and scale height of the disc in the form
\begin{equation}
   \frac{1.1 H}{q^{1/3}R_{p}}+ \frac{1}{q}\frac{50\nu}{\omega R_{p}^{2}}\leq 1.
	\label{eq:gapcrida}
\end{equation}
 Equation (\ref{eq:gapcrida}) gives an estimate of the minimum planet-to-star mass ratio for 
which a planet clears at least $90\%$ of the gas initially in its coorbital region.

In more recent studies, \citet{Fungetal2014} and \citet{Duffell2015} performed numerical
experiments that suggest that the one-sided torque $T_{g}$ due to the 
planet is approximately
\begin{equation}
T_{g}= f_{0}q^{2}\Sigma_{\rm gap} R_{p}^{4}\omega^{2} h^{-3},
\label{eq:fungduffell}
\end{equation}
where $\Sigma_{\rm gap}$ is the surface density in the gap when a steady-state
has been reached and $f_{0}\simeq 0.45\pm 0.543h$ \citep[][and references therein]{Duffell2015}. 
Note that Equation (\ref{eq:fungduffell}) is similar (except for a numerical factor)
to Equation (\ref{eq:H_tot}) in which 
$\Sigma_{0}$ is replaced by $\Sigma_{\rm gap}$. The condition $T_{g}\simeq T_{\nu}$ provides
the surface density in the gap \citep{Fungetal2014,Duffell2015}:
\begin{equation}
\frac{\Sigma_{\rm gap}}{\Sigma_{0}}\simeq \frac{3\pi \nu h^{3}}{f_{0}q^{2}R_{p}^{2}\omega}.
\end{equation}
If our criterion for gap formation is that
$\Sigma_{\rm gap}\lesssim 0.2 \Sigma_{0}$, this implies that a gap forms if
\begin{equation}
q\gtrsim 10\left(\frac{\nu}{\omega R_{p}^{2}}\right)^{1/2} h^{3/2},
\end{equation}
where we have used $f_{0}=0.45$.
We see that the critical value of $q$ for gap formation exhibits a strong dependence on the aspect ratio $h$.

The above criteria for gap opening assume that the planet does not
migrate radially from its initial orbit. Therefore, they are only
valid if the gap opening rate is faster than the radial migration rate of
the planet \citep{Lin1986b,Ward1989}.
For typical circumstellar discs, this condition is satisfied \citep[e.g.,][]{Malik2015}.

\begin{figure}
	\includegraphics[width=\columnwidth]{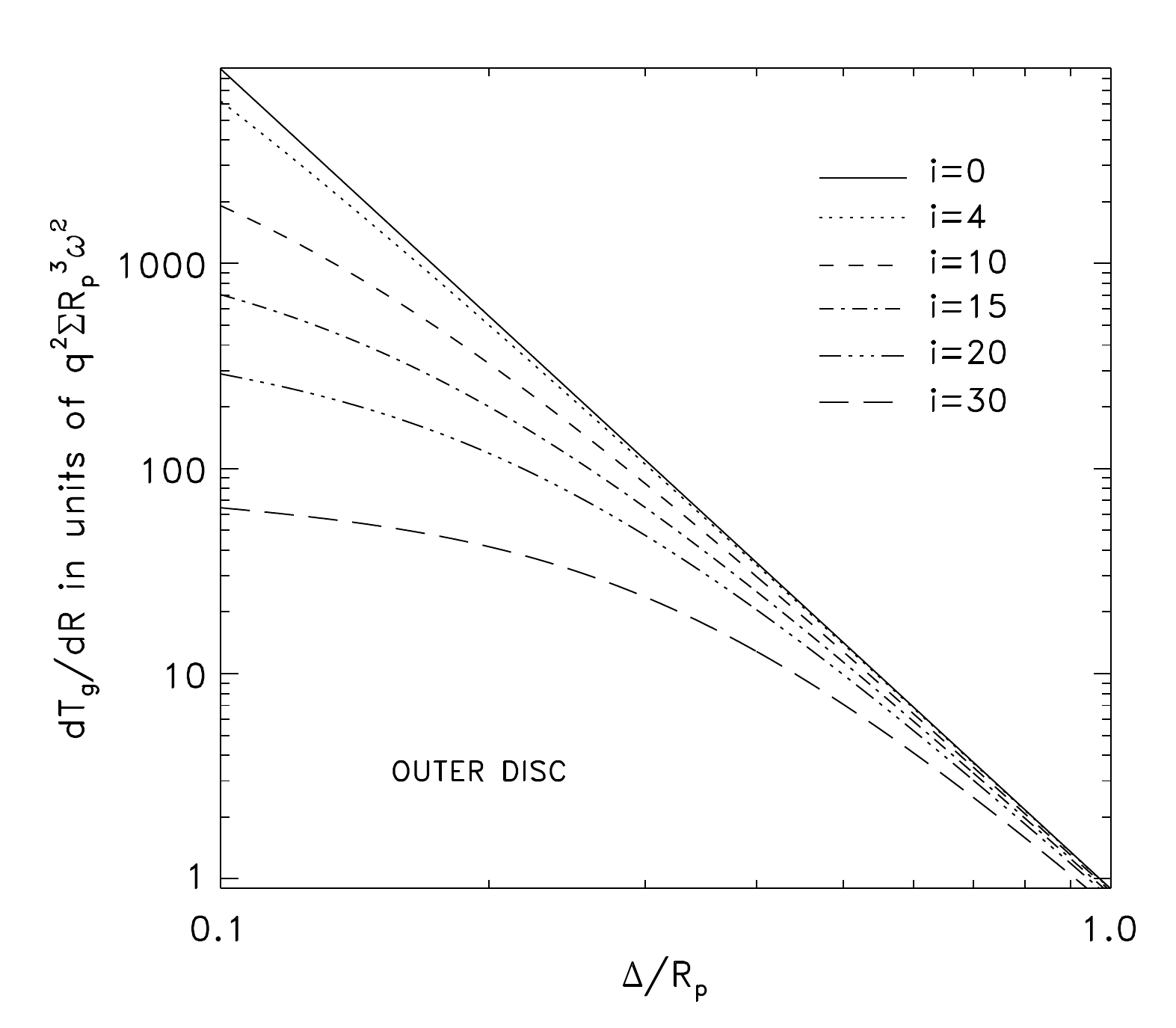}
    \caption{Excitation torque density for $l>0$ and different 
inclinations of the planet's orbit.}
    \label{fig:deltaTg}
\end{figure}

\section{Torques by planets on inclined orbit: the impulse approximation}
 \label{sec:differential_ring}

We consider a thin protoplanetary disc that initially lies in the
plane $z=0$ (hereafter equatorial plane). 
We assume that the planet describes a circular orbit with radius $R_{p}$ and that its orbital plane 
is inclined by an angle $i(t)$ with respect to the midplane of the disc. 
Due to the gravitational interaction of the planet with the disc, tidal torques
lead to a damping of the planetary inclination, implying that $di/dt<0$. 
For realistic protoplanetary discs, the damping timescale $i/|2di/dt|$ is much
larger than the orbital period of the planet. Thus, the planet performs many
orbits before the change in inclination is significant.

We assume the disc to be pressure-less, so that it consists of test
particles, and calculate the disc and planet exchange of angular momentum
as a result of the gravitational interaction of the particles with
the planet. Treating the disc particles as being pressure-less is adequate 
as long as the velocity of the planet relative to the disc particles is 
supersonic \citep[e.g.,][]{Canto2013,Xiang2013}. This condition
is valid for planetary inclinations larger than the disc's aspect ratio\footnote{In fact, 
the relative velocity between the planet and the 
disc particles in the vicinity of the planet is $2 \omega R_{p} \sin (i/2)$, 
and the local Mach number is $2h^{-1}\sin (i/2)$.}. In particular,
for a typical value of $h=0.05$, the planet crosses supersonically 
the disc for inclinations $i\geq 3^{\circ}$.

\subsection{Torques in the impulse approximation}
Without any planet, a certain disc particle will describe circular orbits with radius $R_{d}$
around the central star. In the presence of a planet, the trajectory of
this fluid element will be deflected due to successive gravitational encounters with the planet.
We take the $x$-axis to be in the direction of the ascending line of
nodes of the planet, and take $t=0$ when the planet passes on this
axes, so that its position vector is
\begin{equation}
\vecR_{p}(t)=R_{p}(\cos\phi_{p}, \cos i\sin\phi_{p}, \sin i \sin\phi_{p}),
\end{equation}
where $\phi_{p}=\omega t$ and $\omega=\sqrt{GM_{S}/R_{p}^{3}}$.
The planet reaches its maximum height at the Cartesian points $(0, R_{p}\cos i, R_{p}\sin i)$
and $(0, -R_{p}\cos i,-R_{p}\sin i)$, i.e. at
the azimuthal angles $\pi/2$ and $3\pi/2$. 
The velocity of the planet, $\vecV_{p}$, is
\begin{equation}
\vecV_{p}=\omega R_{p}(-\sin \phi_{p}, \cos i\cos\phi_{p},\sin i\cos\phi_{p}).
\end{equation}

Consider a differential volume element of gas orbiting at a radius $R_{d}$ around the central star.
The angular frequency of this disc particle  is 
$\vecomega=\Omega (R_{d}) \hat{\vece}_{z}$, where $\Omega=\varepsilon\sqrt{GM_{S}/R_{d}^{3}}$,
and $\varepsilon=1$ if the disc rotates counter-clockwise, whereas
 $\varepsilon=-1$ if the disc rotates clockwise.
Note that the planet has a prograde motion respect to the disc if $-\pi/2< i <\pi/2$ and $\varepsilon=1$,
whereas its orbit is retrograde if $-\pi/2< i <\pi/2$ and $\varepsilon=-1$.

The separation vector at the minimum distance between this fluid particle and the planet is
\begin{equation}
\vecd_{\rm min}=\begin{pmatrix} [R_{d}-R_{p}]\cos \phi_{p} \\ [R_{d}-R_{p}\cos i]\sin \phi_{p} \\  -R_{p}\sin i\sin\phi_{p} \end{pmatrix}.
\label{eq:dist_min}
\end{equation}
Its modulus is
\begin{equation}
d_{\rm min}^{2}=\left[\Delta^{2}+4 R_{d}R_{p}\sin^{2}(i/2)\sin^{2}\phi_{p}\right]^{1/2},
\end{equation}
where $\Delta\equiv R_{d}-R_{p}$.

For streamlines passing close enough to the perturber, $d_{\rm min}\ll R_{p}$ (which requires
that $\sin i\ll 1$), and the relative velocity between the disc particle and the planet is
\begin{equation}
\vecv_{\rm rel}=R_{p}\begin{pmatrix} [\omega -\Omega]\sin\phi_{p} \\ [\Omega -\omega \cos i] \cos \phi_{p} \\  -\omega \sin i  \cos\phi_{p}\end{pmatrix}.
\label{eq:vecv_rel}
\end{equation}
In the impulse approximation, we assume that the close encounter between the disc particle and
the perturber occurs with impact parameter $d_{\rm min}$ and velocity $v_{\rm rel}$, and that the
deflection angle $\delta_{e}$ is small enough so that the trajectory of the disc particle
is approximately rectilinear. In the planet frame, the deflection angle
is 
\begin{equation}
\cot^{2}\left(\frac{\delta_{e}}{2}\right)=\frac{v_{\rm rel}^{4}d_{\rm min}^{2}}{G^{2}M_{p}^{2}},
\label{eq:cotdeltae}
\end{equation}
where we recall that $M_{p}$ is the mass of the planet.
From Equation (\ref{eq:vecv_rel}), we have
\begin{equation}
v_{\rm rel}^{2}=R_{p}^{2}\left[(\Omega-\omega)^{2}+4\omega \Omega \sin^{2}(i/2)\cos^{2}\phi_{p}
\right].
\label{eq:vrel_modulus}
\end{equation}
The velocity of the fluid element immediately after one gravitational scattering, in the system
of reference of a nonrotating observer, is
\begin{equation}
\vecV_{f}={\mathcal{R}}\vecv_{\rm rel}+\vecV_{p},
\end{equation}
where ${\mathcal{R}}$ is the rotation matrix of angle $\delta_{e}$ around the axis parallel to the vector $\vecd_{\rm min}\times \vecv_{\rm rel}$.  

The disc particle remains orbiting in the $z=0$ plane only if
$\vecd_{\rm min}\times \vecv_{\rm rel}$ is parallel to the $z$-axis,
which occurs when $i=0$. In a general case, disc particles may be scattered to a tilted plane.  
After one encounter, the specific (orbital) angular momentum
of a disc particle $\vecL$ will change from its unperturbed value $\vecL_{i}$
to a value $\vecL_{f}$. A change in the direction of
the angular momentum corresponds to a warp, whereas a change in the
magnitude of $\vecL$ corresponds to a change in $R_{d}$. In the coplanar case ($i=0$), $\vecL$
is always parallel to $\hat{\vece}_{z}$ and the planetary torques lead to a redistribution
of the mass in the plane of the disc.
In those encounters for which $\vecL_{f}$ and $\vecL_{i}$ have different directions but the same magnitude, the fluid element is scattered to another plane but will have the same orbital radius.  Fluid elements will move radially outwards or inwards from the planet's position when $\vecL$ changes its magnitude during the collision.  
Since we are interested in the radial redistribution of the mass in the disc,
our aim is to calculate the rate at which $L_{f}-L_{i}\equiv |\vecL_{f}|-|\vecL_{i}|$ changes due to 
successive encounters with the planet.

Initially, the specific angular momentum of a fluid element about the central star is:
\begin{equation}
\vecL_{i}=\Omega R_{d}^{2}\hat{\vece}_{z}.
\end{equation}
After the gravitational deflection with the planet, the specific angular momentum is
\begin{equation}
\vecL_{f}=\vecR_{d}\times ({\mathcal{R}}\vecv_{\rm rel}+\vecV_{p}).
\end{equation}
where $\vecR_{d}=\vecR_{p}+\vecd_{\rm min}$.
The change in the magnitude of the angular momentum can be written
in terms of ${\mathcal{R}}_{1}\equiv {\mathcal{R}}-{\mathcal{I}}$,
where ${\mathcal{I}}$ is the identity matrix, as 
\begin{eqnarray}
L_{f}-L_{i} \simeq \varepsilon \hat{\vece}_{z}\cdot [\vecR_{d}\times ({\mathcal{R}}_{1}\vecv_{\rm rel})]
=\varepsilon R_{d} ({\mathcal{R}}_{1} \vecv_{\rm rel})\cdot \vece_{\phi},
\label{eq:lf_minus_li}
\end{eqnarray}
where $\vece_{\phi}$ is the unitary vector in the azimuthal direction of the particles during
the encounter: $\vece_{\phi}=(-\sin\phi_{p},\cos\phi_{p},0)$.

The (gravitational) torque acting upon an elementary ring of radius
$R$ and width $\delta R$ is 
\begin{equation}
\delta T_{g} (R) =\delta R\int_{0}^{2\pi} \varepsilon (L_{f}-L_{i})\Sigma v_{\rm rel} 
\frac{d\phi_{p}}{2\pi}.
\label{eq:deltaTg_prev}
\end{equation}

Hereafter we specialize in the prograde case ($\varepsilon=1$) but the retrograde case does not
pose any additional complication. 
Substituting Eqs (\ref{eq:vrel_modulus}) and (\ref{eq:lf_minus_li}) into Eq.~(\ref{eq:deltaTg_prev}), 
the radial torque density in the impulse approximation is
\begin{equation}
\begin{aligned}
&\frac{d T_{g}(l)}{dR}=\frac{\Sigma (R) \omega^{2} R_{p}^{3}}{2\pi} \times\\
 &\int_{0}^{2\pi}[( {\mathcal{R}}_{1} \tilde{\vecv}_{\rm rel})\cdot \vece_{\phi}]
\left(\frac{9}{4}l^{2}+2 \left(2-3l\right)\sin^{2}(i/2)\cos^{2} \phi_{p}\right)^{1/2} d\phi_{p},
 \label{eq:delta_inc}
 \end{aligned}
\end{equation}
where we have introduced the dimensionless variables $l\equiv (R-R_{p})/R_{p}$ and
$\tilde{\vecv}_{\rm rel}\equiv \vecv_{\rm rel}/(\omega R_{p})$. 
In the Appendix, we show that the expression for $T_{g}$ 
derived by \citet{LinPapaloizou1979} given in Equation
(\ref{eq:Tg_impulse_approx_coplanar}) is recovered for $i=0$.
In order to derive Eq.~(\ref{eq:delta_inc}), we have assumed small angle inclinations.
Therefore, our approximation is inaccurate for
large inclinations.

Figure \ref{fig:deltaTg} shows the radial torque density $dT_{g}/dR$ on the outer disc 
(note that $l>0$ implies that the ring lies in the outer
disc). As expected, the torque on a given ring decays for larger inclination angles of 
the planetary orbit. We see that as $i$ increases, the profile of $dT_{g}/dR$ vs 
$l$ flattens at low $l$.

In Section \ref{sec:cop_case} we mentioned that the impulse
approximation in the coplanar case accounts for the scalings of $T_{g}$
and gives its magnitude {\em to within a factor of $2$}.
It is convenient to introduce a constant factor
of the order of the unity, $\xi$, such that the corrected gravitational torque density
$dT_{g}^{\rm cor}/dR$ becomes
\begin{equation}
\frac{dT_{g}^{\rm cor}}{dR}= \xi \frac{dT_{g}}{dR}.
\end{equation} 
By comparing with numerical simulations, we will verify whether
the impulse approximation also predicts the correct scaling for inclined
planetary orbits, and if it does, we will determine the value of $\xi$
that best matches the simulation results.

\subsection{The minimum and maximum values of $l$}
\label{sec:cutoffs}
The radial torque density $dT_{g}/dR$ derived in the last section is not 
valid for $|l|<l_{\rm min}$ either for $|l|>l_{\rm max}$, where $l_{\rm min}$
is the minimum impact parameter for which the assumptions of small deflection and 
null thickness of the disc are still valid, and $l_{\rm max}< 1$ because the impulse 
approximation breaks down for encounters with large impact parameters  
as the orbits cannot be assumed rectilinear.
For our purposes, the exact value of $l_{\rm max}$ is not relevant because
the torque density decays very quickly with $|l|$, so we will take $l_{\rm max}=1$.
The value for $l_{\rm min}$ is a more delicate issue and it is discussed in the following.

As we have ignored the thickness of the disc in our derivation, $l_{\rm min}$ should
be comparable to or larger than $h$. In addition, the deflection angle in encounters with
an impact parameter of $l_{\rm min}$ should be small.

In the coplanar case, the deflections
are large in the coorbital region, i.e. at distances $\sim R_{H}$ from the
planet. For planets in inclined orbits, the relative velocity between the
planet and the disc particles in the vicinity of the planet is $2 \omega R_{p} \sin (i/2)$.
Since the deflection angle decreases with the relative velocity (see Eq.~\ref{eq:cotdeltae}), the
condition of small deflections could be fulfilled for impact parameters 
less than $R_{H}$ for inclined planets. For illustration, consider the
extreme case in which the orbit of the planet is coplanar to the disc but moves
in a retrograde orbit. In this case, 
deflections are only large within the planetary accretion radius $r_{\rm acc}$ defined as
$2GM_{p}/V_{\rm rel}^{2}$. 
Since the relative velocity in this case is $2\omega R_{p}$, we obtain
that $r_{\rm acc}/R_{H}=0.7q^{2/3}$, which implies that $r_{\rm acc}$ is 
a factor of $100$ smaller than $R_{H}$ for $q=10^{-3}$. In general, we can state that if
$r_{\rm acc}\ll R_{H}$ or, equivalently, when $i\gg i_{\rm crit}\equiv 2\arcsin(0.85 q^{1/3})$, the
minimum impact parameter is given by $r_{\rm acc}$ and not by the Hill radius, 
as there is no coorbital region at all.

For the values of $i$, $q$ and $h$ explored in this paper, 
the values for $R_{H}$, $H$ and $r_{\rm acc}$ are all of the same magnitude within a factor of $2$.
In view of this, we take $l_{\rm min}={\rm max}\{(q/3)^{1/3},h\}$, unless otherwise
stated. Finally, we assume that the gravitational torque density is null (i.e. $dT_{g}/dR=0$) 
at $|l|<l_{\rm min}$ and at $|l|>l_{\rm max}$. This simple cutoff in the torque density
is commonly adopted in the coplanar case \citep[e.g.][]{Cridaetal2006,Kanagawaetal2015}.

\section{Gaps by planets on inclined orbits}
\label{sec:gapsmethods}
\subsection{Steady-state gaps by planets in fixed orbits}
\subsubsection{Viscous criterion for the formation of a deep gap by planets in fixed orbits}
\label{sec:criterion_i}
It is possible to derive a criterion for gap formation similar to that
of Equation (\ref{eq:criteria_LP93})  but for a planet that is forced to move on an
orbit with non-zero inclination, i.e. ignoring the damping of inclination.
Such a criterion is useful if the damping timescale is much larger than the timescale
for opening the gap. This situation may occur when the planet has acquired
its inclination after the gas is well depleted, or if the relative
inclination of the planet with the disc is maintained by some 
external source such as accretion of mass (which may change the orientation 
of the disc) or through resonant inclination excitation by a second giant planet
\citep{Thommes2003}.  
This criterion may be also useful to 
interpret simulations that are started after a stage where the disc is evolved with
the planet on a fixed inclined orbit \citep[e.g.,][]{Bitsch2013}.

As in the coplanar case to derive the gap opening criterion (see \S \ref{sec:cop_case}),
we need to calculate the one-sided gravitational torque. 
To derive a gap criterion, we estimate the torque on the external disc by 
assuming that $\xi=1$, $l_{\rm max}=1$ and $l_{\rm min}=q^{1/3}$ (see \S \ref{sec:cutoffs}
and Lin \& Papaloizou 1979).
Having fixed $l_{\rm min}$ and $l_{\rm max}$
we can compute numerically  
the total torque acting on the external side of the disc given by
\begin{equation}
T_{g}(q,i)=R_{p}\int_{l_{\rm min}}^{l_{\rm max}} \frac{dT_{g}}{dR} dl,
\label{eq:Tg_ext}
\end{equation}
as a function of $q$ and $i$, using 
Equation (\ref{eq:delta_inc}).
For $0\leq i\leq 30^{\circ}$ and $5\times 10^{-5}\leq q\leq 2\times 10^{-2}$, 
we provide an empirical fit of the resultant $T_{g}(q,i)$ with an error less than $12\%$.
The criterion condition is derived by imposing 
that a gap forms if $T_{g}\geq T_{\nu}$, where $T_{\nu}$ is given in Equation (\ref{eq:Tensor_visc}).
In the following, we write the gap criterion in terms of $q_{\textsc{\tiny -3}}\equiv q/10^{-3}$
and $i_{\textsc{\tiny 10}}\equiv i/10$ ($i$ in degrees). A deep gap is predicted to form if:
\begin{equation}
\tilde{C}(q,i) q\geq \frac{32\nu}{\omega R_{p}^{2}},
 \label{eq:i10}
\end{equation}
where
\begin{equation}
\tilde{C}(q,i) = \left\{
        \begin{array}{ll}
            \frac{q_{\textsc{\tiny -3}}}{(1+i_{10}^{3.5})(q_{\textsc{\tiny -3}}+
0.18i_{\textsc{\tiny 10}}^{2})^{\beta (i)}} & \quad i < 17^{\circ}, \\[0.5ex]\\
            0.22 q_{\textsc{\tiny -3}} \exp\left[-\Psi (q,i)-\Pi (i)\right] & \quad  i > 17^{\circ},
        \end{array}
    \right.
\end{equation}
and
\begin{equation}
\beta(i)=1-0.26i_{\textsc{\tiny 10}},
\end{equation}
\begin{equation}
\Psi(q,i)=\left(\frac{q_{\textsc{\tiny -3}}}{1.3i_{\textsc{\tiny 10}}+0.2}\right)^{0.34},
\end{equation}
and
\begin{equation}
\Pi(i)=\frac{1}{2}(i_{\textsc{\tiny 10}}-1.7)^{2}.
\label{eq:fu_Pi}
\end{equation}

In the particular case $i=0$, $\tilde{C}(q,0)=1$, and
the well-known viscosity criterion $q\geq 32\nu/(\omega R_{p}^{2})$ is recovered [see, e.g.,
Equation (23) in \citet{Lin1993}].
Given the disc viscosity and the inclination $i$
of the planet, we can obtain the value of $q_{\rm crit}$ for the gap opening in the surface density. 
For a typical value of the effective viscosity of $10^{-5}\omega R_{p}^{2}$ and for 
$i=0, 10^{\circ}$, $20^{\circ}$ and $30^{\circ}$, we find $q_{\rm crit}=0.5\times 10^{-3}$,
$0.8\times 10^{-3}$, $1.8\times 10^{-3}$ and $3.0\times 10^{-3}$, respectively. 
We expect that for values of $q$ larger than $q_{\rm crit}$, the surface density at the bottom
of the gap should be $\leq 0.25\Sigma_{0}$ (see \S \ref{sec:cop_case}).
In Section \ref{sec:num_sim}, we present numerical experiments to test whether this prediction
is correct.

For a planet that undergoes inclination damping, we define $i_{\rm open}$ as
the inclination of the planet's orbit at the time at which the surface density at the gap is 
$\sim 0.2\Sigma_{0}$. If we give the value of $i_{\rm open}$, then Equations (\ref{eq:i10})-(\ref{eq:fu_Pi})
provide a lower limit for $q$, just by replacing $i$ for $i_{\rm open}$.

\begin{figure*}
\centering
 \includegraphics[width=1\textwidth]{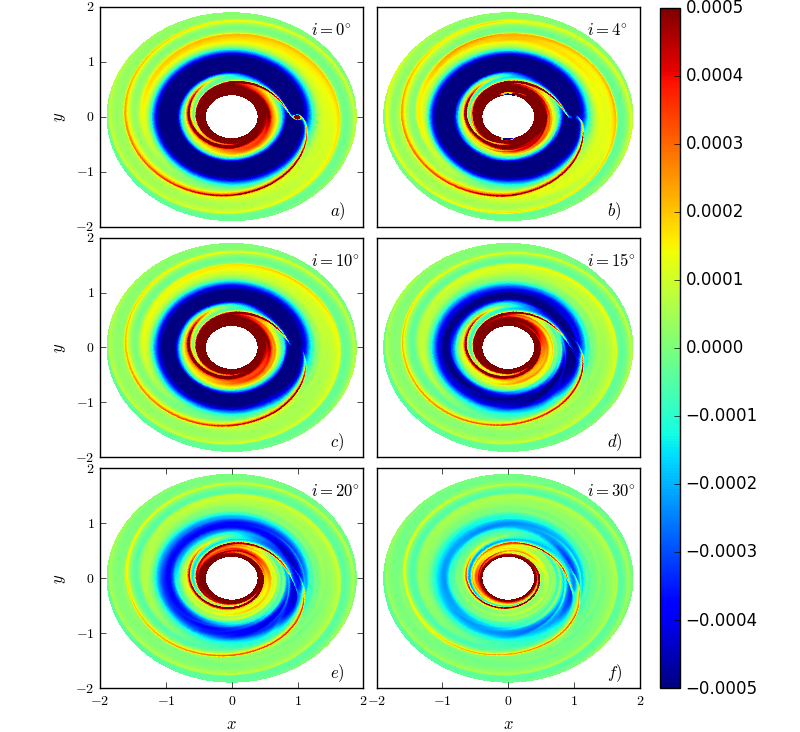}
 \caption{Perturbation of volume density $(\rho-\rho_i)$ at $z=0$ after $t=200$ orbits, for different inclinations.
In all cases, $q=10^{-3}$, $h=0.05$, and $\nu=10^{-5}$ (Runs 1 to 6). 
In all the plots, the planet is at $x=\cos i$, $y=0$ and $z=\sin i$, i.e.~it is at its 
maximum height from the disc. Note that the scale is linear (not logarithmic).}
\label{fig:disc_integer}
\end{figure*}

\begin{figure*}
\centering
 \includegraphics[width=1\textwidth]{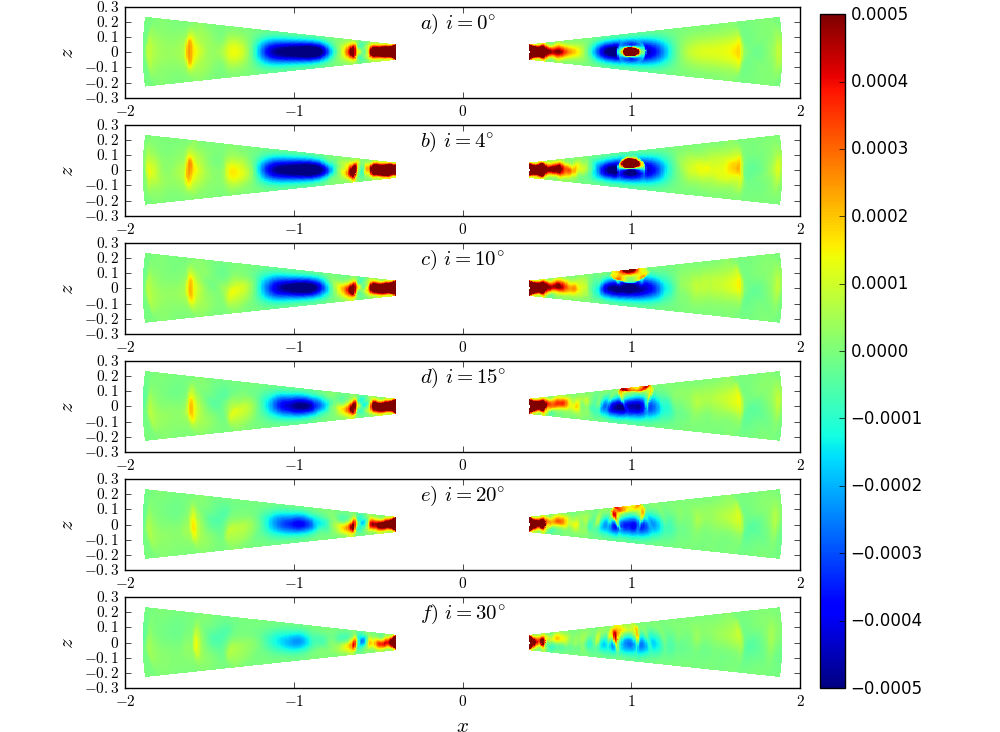}
 \caption{Same as Figure \ref{fig:disc_integer}  but along vertical cross sections in the plane $y=0$.}
\label{fig:disc1}
\end{figure*}

\begin{figure*}
\centering
 \includegraphics[width=\columnwidth]{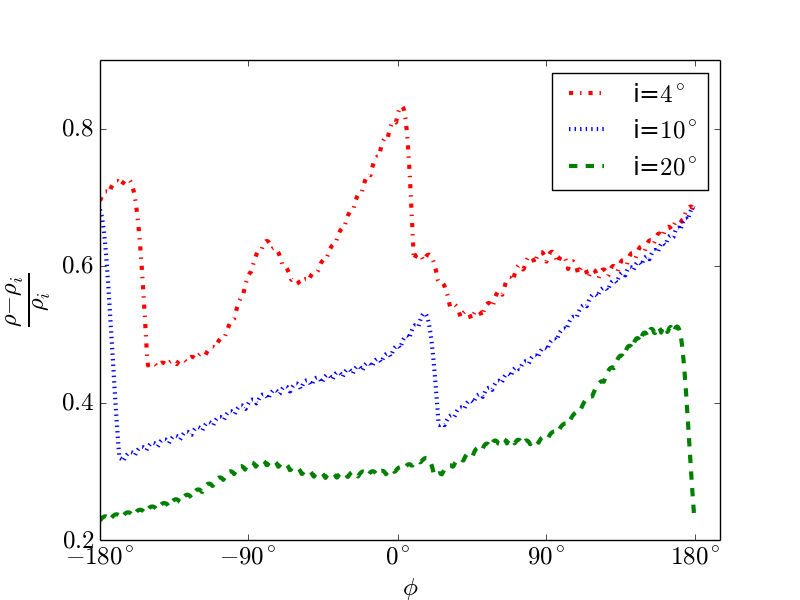}
 \caption{Perturbation of volume density along the crest of the outer spiral wave, at $z=0$.}     
 \label{fig:wake}
\end{figure*}

\subsubsection{Stationary gap profile in the approximation of local deposition of the torque}
\label{sec:method1}

Under the assumption that the gravitational torque is locally
(instantaneously) deposited in the disc (\emph{i. e. } ignoring
the propagation of waves before damping)
a steady state is reached when the gravity is balanced by the viscous torque at every ring of the disc 
\citep[see, for instance,][]{Varniereetal2004, Cridaetal2006}. 
The radial densities of the viscous torque, $dT_{\nu}/dR$, and
of the gravitational torque, 
$dT_g/dR$ can be obtained from Equations (\ref{eq:Tensor_visc})
and  (\ref{eq:delta_inc}). Then, equating these torque densities, we obtain
a differential equation that describes the gap structure: 
\begin{equation}
   \frac{1}{\Sigma}\frac{d\Sigma}{dR} =\frac{\xi}{3\pi\nu R^2\Omega\Sigma}\frac{dT_g}{dR}-\frac{1}{2R}.
	\label{eq:Sigma_diff}
\end{equation}
To solve Equation (\ref{eq:Sigma_diff}), we need to choose the boundary condition.
In the outer parts of the disc, far away from the planet, we expect that the surface density
remains essentially unperturbed. Thus, it is ordinary to integrate Equation (\ref{eq:Sigma_diff})
from a certain point $R_{\rm max}\gg R_{p}$ inwards down to $R_{p}+r_{\rm min}$, where
$r_{\rm min}$ is the distance from the planet where the impulse approximation breaks down.
We may continue the integration of the equation in the inner disc by adopting a reasonable value
for $\Sigma$ at $R_{p}-r_{\rm min}$. For instance, \citet{Cridaetal2006} assumed that
$\Sigma(R)\propto R^{-1/2}$ between $R_{p}-r_{\rm min}$ and $R_{p}+r_{\rm min}$, and
adopted $r_{\rm min}=2R_{H}$, with $R_{H}$ the Hill radius.

\begin{figure*}
\centering
 \includegraphics[scale=0.55]{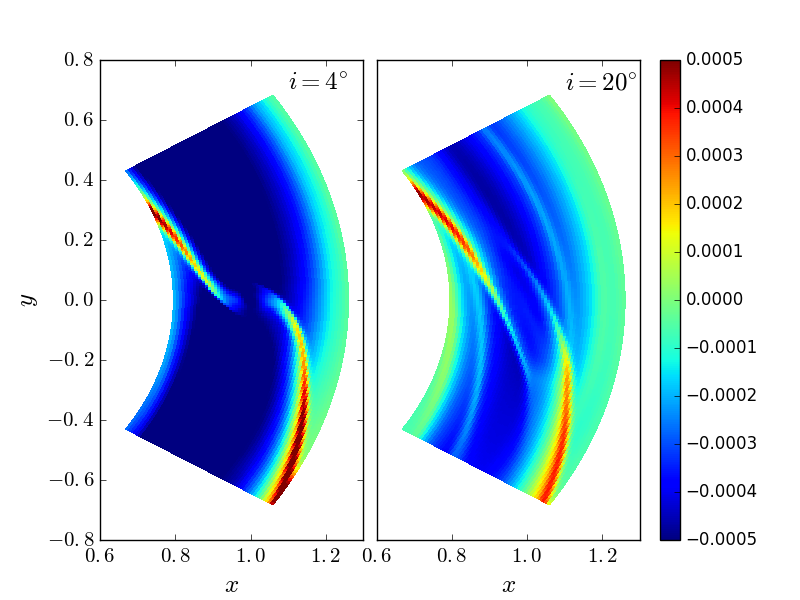}
 \caption{Zoom of the perturbed density $\rho-\rho_{i}$ at $z=0$ and $t=200$ orbits,
for inclinations $i=4^{\circ}$ (Run 2) and $i=20^{\circ}$ (Run 5). 
The planet is at $x=0.9975$, $y=0$ and $z=0.0697$ in the left panel and 
at $x=0.9396$, $y=0$ and $z=0.3420$ in the right panel.}     
 \label{fig:disk_zoom}
\end{figure*}

\begin{figure}
	\includegraphics[width=\columnwidth]{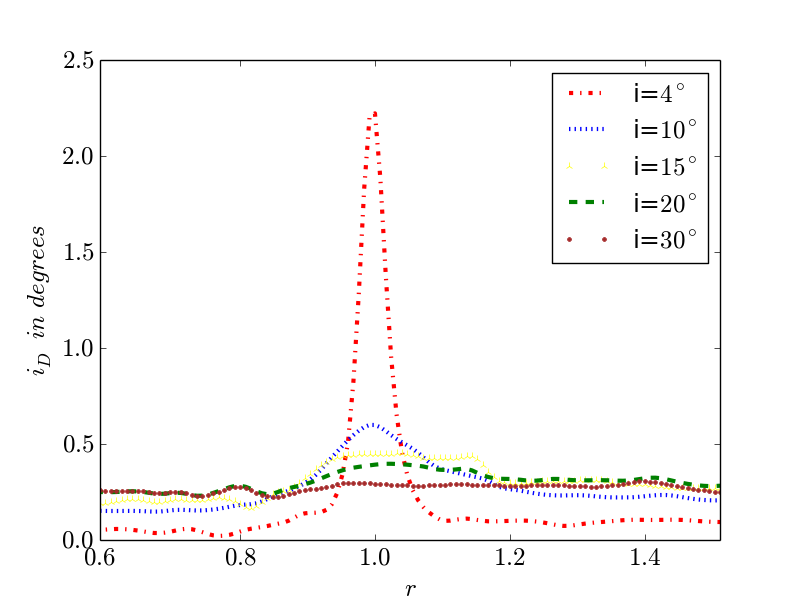}
    \caption{Inclination angle of the disc, $i_{D}$, as a function of $r$ after $200$ orbits, 
for $q=10^{-3}$, $h=0.05$ and different $i$. The rms in the measurement of $i_{D}$
is $0.05^{\circ}$. Thus, $i_{D}$ values below $\sim 0.05^{\circ}$ are not significant.}
    \label{fig:inclination}
\end{figure}

In the coplanar case, the resultant gap profile using the
instantaneous damping approximation (i.e. using Eq.~\ref{eq:Sigma_diff}) 
has been extensively studied.  It was found that the predicted gap profile is
consistent with the simulated gaps only for high disc viscosities. At
lower viscosities, the predicted gaps are wider than those
observed in numerical simulations \citep[for instance,
see][]{Varniereetal2004, Cridaetal2006}.  Moreover, at these low
viscosities the predicted scaling relation between the surface density
averaged over the bottom of the gap ($\Sigma_{\rm gap}$) and $q$ is
also incorrect; it cannot explain why $\Sigma_{\rm gap}$ scales as a
power-law with $q$, as found in numerical simulations
\citep[e.g.,][]{Duffelletal2014, Fungetal2014}.

\subsubsection{Gap depth in a zero-dimensional analysis}
\label{sec:method2}

In order to reproduce the dependence of gap depth on $q$, viscosity and
$h$ observed in hydrodynamical simulations for the coplanar case,
\citet{Fungetal2014}, \citet{Kanagawaetal2015} and \citet{Duffell2015} have invoked a
``zero-dimensional''  approximation, which assumes that the torque occurs only within the
width of the gap.
Under this approximation, the total one-sided torque can be written as
\begin{equation}
T_{g}=\hat{f}_{0}(i,h) q^{2}\Sigma_{\rm gap} R_{p}^{4}\omega^{2}  h^{-3}.
\end{equation}
By choosing reasonable values for $l_{\rm min}$ and $l_{\rm max}$,
the scaling of the prefactor $\hat{f}_{0}$ with $i$ and $h$ can be computed 
from Equations (\ref{eq:delta_inc}) and (\ref{eq:Tg_ext}). 
As discussed in \S\ref{sec:cutoffs}, we take $l_{\rm min}={\rm max}\{(q/3)^{1/3},h\}$ and $l_{\rm max}=1$. 
For a fixed aspect ratio $h_{0}$, we can find 
how $\hat{f}_{0}$ depends on the inclination $i$. To do so,
we have evaluated the integral given in Equation (\ref{eq:Tg_ext})
for $l_{\rm min}=1.1h$, $l_{\rm max}=1$, and different inclinations. 
For the particular value of $h=0.05$, we find that $\hat{f}_{0}(i,h=0.05)$ can be fitted as
\begin{equation}
\begin{aligned}
\hat{f}_{0}(i,h=0.05)=&0.744\exp \left(-\frac{i}{3}\right)-0.450\exp\left(-0.7i\right)\\
&+0.155\exp\left(-\frac{i^{0.9}}{5.6}\right)+0.0004,
\end{aligned}
\label{eq:hatfo}
\end{equation}
where $i$ is the inclination angle in degrees. This fit is valid for $i<35^{\circ}$, with a fractional
error lower than $4\%$.  For the calibration of the magnitude of $\hat{f}_{0}$ (i.e., to fix the value of $\xi$), 
we have used the condition that $\hat{f}_{0}=0.45$ for $i=0$ (see \S \ref{sec:cop_case}). 
It is worth noting that $\hat{f}_{0}$ decays a factor of $100$ between
$i=0$ and $i=30^{\circ}$. 

Once $\hat{f}_{0}$ is determined, Duffell's model predicts $\Sigma_{\rm gap}$
through the formula
\begin{equation}
\frac{\Sigma_{\rm gap}}{\Sigma_{0}}=
\left(1+\frac{\hat{f}_{0} q^{2}R_{p}^{2}\omega}{3\pi\nu h^{3}}\right)^{-1}.
\label{eq:duffell_eq}
\end{equation}
\citet{Duffell2015} also provides a recipe to obtain the profile of the gap.
However, the calculation of the profile requires knowledge of the angular momentum
flux due to the damping of the planetary wake, which is uncertain for planets
in inclined orbits. Therefore, we only use the ``zero-dimensional analysis'' to predict the 
gap depth.

\subsection{The formation of the gap in the local approximation: time evolution
and inclination damping}
\label{sec:timevolution}
The steady-state gap formed by a coplanar planet has been studied in great detail 
because the timescale to reach the steady state is
shorter than the migration timescale (see \S \ref{sec:cop_case}). For a planet on
an inclined orbit, the inclination damping timescale may be comparable to or smaller
than the timescale for gap opening if the disc is sufficiently massive \citep{MarzariNelson2009,Xiang2013,Bitsch2013}. 
Under these circumstances, it is necessary to consider
the time evolution of the disc surface density in order to include the 
dependence of the planet's inclination with time. 

Suppose that $i(t)$ is known. Then, the torque density $dT_{g}/dR$ depends implicitly on time
through $i(t)$. Assuming that the evolution
of the disc is axisymmetric, $\Sigma (R,t)$ can be computed by solving the continuity equation
\begin{equation}
\frac{\partial\Sigma}{\partial t}+\frac{1}{R} \frac{\partial}{\partial R}(R\Sigma v_{\mbox{{\tiny $R$}}})=0,
\end{equation}
the radial momentum equation 
\begin{equation}
\frac{\partial v_{\mbox{\tiny $R$}}}{\partial t}+v_{\mbox{\tiny $R$}}\frac{\partial v_{\mbox{\tiny $R$}}}{\partial R} =-\frac{GM_{S}}{R^{2}}+\frac{v_{\phi}^{2}}{R}
-\frac{1}{\Sigma}\frac{\partial}{\partial R}(\Sigma c_{s}^{2}),
\end{equation}
and the conservation of angular momentum
\begin{equation}
\frac{\partial L}{\partial t}+\frac{1}{R}\frac{\partial}{\partial R}(Rv_{\mbox{\tiny $R$}}L)=
\frac{\nu}{R}\frac{\partial}{\partial R}(\Sigma R^{3}\Omega')+\frac{\xi}{2\pi R}
\frac{dT_{g}}{dR},
\label{eq:momentum}
\end{equation}
where  $\Omega\equiv v_{\phi}/R$, $\Omega'\equiv d\Omega/dR$ and 
$L\equiv \Sigma\Omega R^{2}$ is the angular momentum of a differential ring
in the disc (e.g., Pringle 1981). In Equation (\ref{eq:momentum}), we have employed
the local deposition approximation.

As a particular case, one can derive the
time evolution of $\Sigma(R,t)$ in the presence of a planet on a fixed orbit,
i.e. $i(t)=i_{0}=$const, in a disc that initially has no gap.
To test whether the present 1D model is successful or not, it is sufficient to check if
$\Sigma(R,t)$ obtained in full 3D hydrodynamical simulations is correctly reproduced for any value of $i_{0}$. 
If so, the formalism should also satisfactorily predict $\Sigma(R,t)$ for an arbitrary function $i(t)$.

\begin{figure}
	\includegraphics[width=\columnwidth]{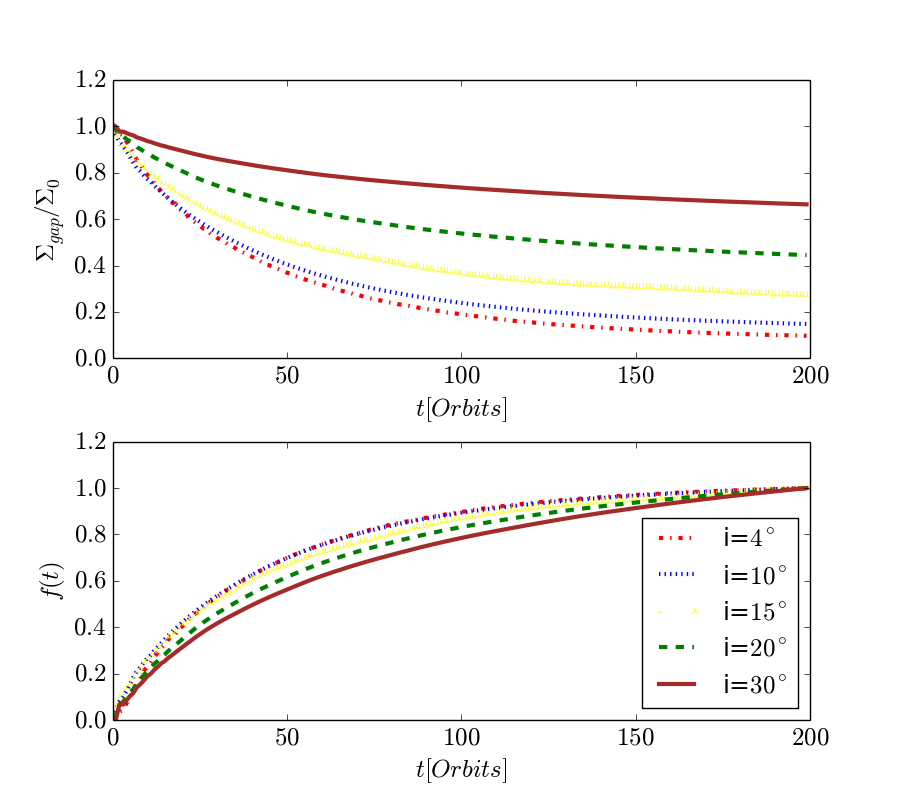}
    \caption{Temporal evolution of $\Sigma_{\rm gap}$ for different inclinations. In all cases
$q=10^{-3}$ and $h=0.05$.}
    \label{fig:gapsito}
\end{figure}

\begin{figure}
	\includegraphics[width=\columnwidth]{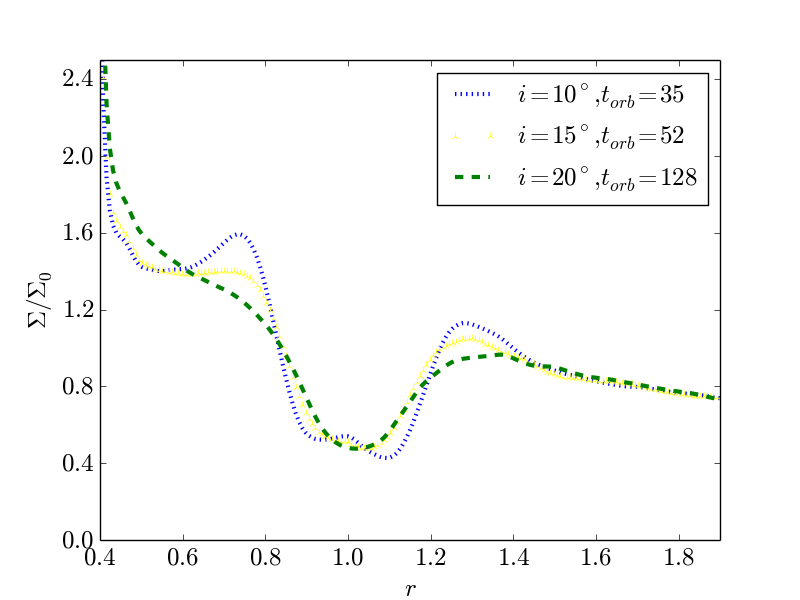}
    \caption{Radial profiles of the azimuthally averaged surface density $\Sigma$ at the time when
$\Sigma_{\rm gap}=\Sigma_{0}/2$. The corresponding time depends on inclination $i$ and is
quoted at the corner of the figure.} 
    \label{fig:S2}
\end{figure}

\begin{figure*}
\centering
    \includegraphics[scale=0.9]{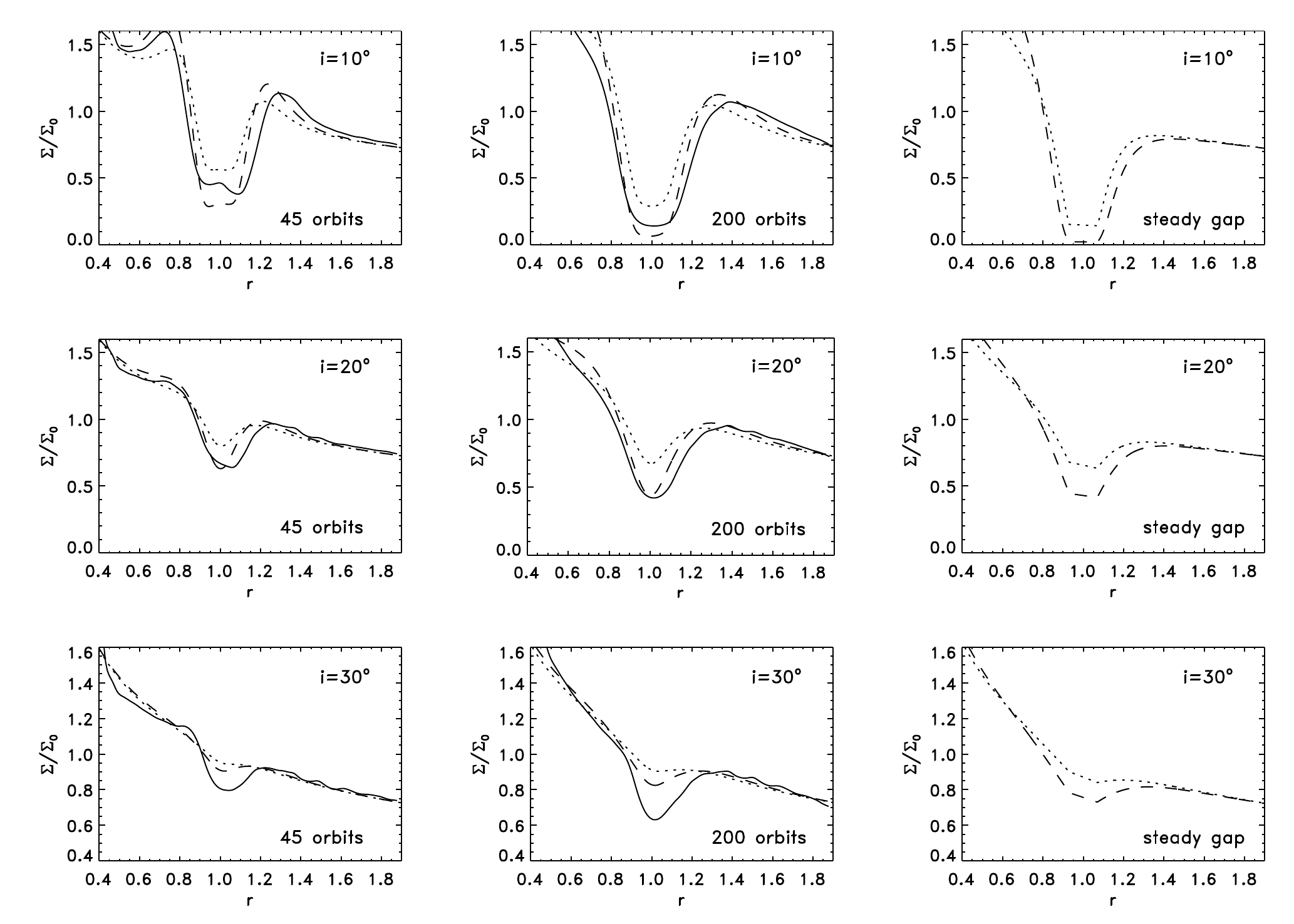}
 \caption{ Gap profiles using the local damping approximation (see \S \ref{sec:timevolution}) 
with $\xi=1$ (dotted lines) and $\xi=2$ (dashed lines) together with
those from numerical simulations (solid lines) after $45$ and $200$ orbits, for different inclinations.
The stationary gap profiles using the local damping approximation are displayed in the right column.
In all cases $q=10^{-3}$ and $h=0.05$.}     
 \label{fig:sigma1}
 \end{figure*}

\begin{figure*}
\centering
    \includegraphics[scale=0.85]{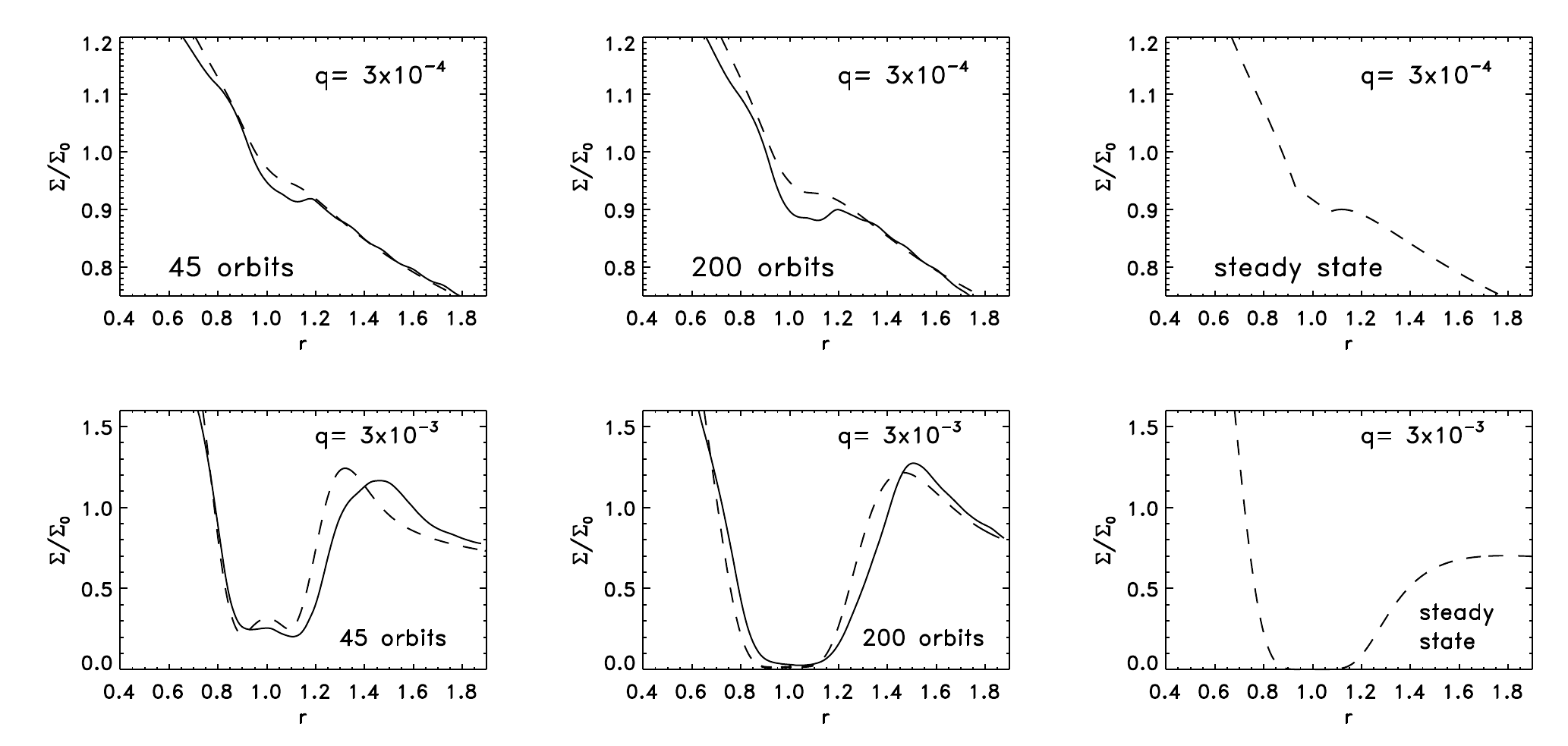}
 \caption{ Comparison between the gap profiles in the simulations with $i=20^{\circ}$ 
and different $q$ (solid lines) with those
using the local damping approximation with $\xi=2$ (dashed lines). In all cases $h=0.05$.}     
 \label{fig:diffm}
 \end{figure*}

\begin{table}
	\centering
	\caption{Parameters of the simulations. $q_{\rm crit}$ is the viscous critical value for
gap formation according to Eqs (\ref{eq:i10})-(\ref{eq:fu_Pi}). 
In all the cases, the kinematic viscosity is $10^{-5}$. Remind that $q$ denotes the planet to star mass ratio. }
	\label{tab:simulations}
	\begin{tabular}{cccccc} 
              Run  &  $i$ & $h$ & $q_{\rm crit}/10^{-3}$ & $q/10^{-3}$ &  $\Sigma_{\rm gap}/\Sigma_0$\\
                    &    deg   &   &   &  & at $200$ orbits\\
		\hline
		1 & 0 & 0.05 &  $0.5$ &$1$ & 0.093\\
		2 & 4 & 0.05 &  $0.55$ & $1$& 0.097\\
		3 & 10 & 0.05 & $0.8$ & $1$ & 0.148\\
		4 & 15 & 0.05 &  $1.2$ & $1$& 0.273\\
		5 & 20 & 0.05 & $1.8$ & $1$ & 0.445\\
		6 & 30 & 0.05 & $3.0$ & $1$ & 0.664\\
		7 & 15 & 0.05 & $1.2$ & $3$ & 0.011\\
		8 & 20 & 0.05 & $1.8$ & $3$ & 0.037\\
		9 & 20 & 0.05 & $1.8$ &$0.3$ & 0.927\\
		10 & 20 & 0.025 & $1.8$ & $1$ & 0.264\\
		\hline
	\end{tabular}
\end{table}

\section{NUMERICAL SIMULATIONS}
 \label{sec:num_sim}
\subsection{The code and initial conditions}
\label{sec:fargo3d} 
In our study of the gap opening by a planet on inclined orbit, we use a spherical coordinate system $(r,\theta,\phi)$, 
where $r$ is the radial coordinate, $\theta$ is the polar angle, and $\phi$ is the azimuthal angle.
The hydrodynamical equations describing the flow are the equation of continuity
\begin{equation}
   \frac{\partial\rho}{\partial t}+ \vecna\cdot (\rho\vecv)=0
	\label{eq:hydrodynamics}
\end{equation}
and momentum equation
\begin{equation}
   \frac{\partial \vecv}{\partial t}+ (\vecv\cdot\nabla)\vecv=-\frac{1}{\rho}\vecna P-\vecna\Phi+\mathbf{f_{\nu}}.
	\label{eq:hydrodynamics1}
\end{equation}
Here $\rho$ is the density, $\vecv$ the velocity, $\Phi$ is the gravitational potential
and  $\mathbf{f_\nu}$ represents the viscous force per unit volume. 
The disc is assumed to be locally isothermal, i.e. the pressure is given by
\begin{equation}
       P=\rho c_s^2,
	\label{eq:pressure}
\end{equation}
where $c_s(r)$ is the isothermal sound velocity.

We use the hydrodynamic FARGO3D code \citep{Benitez-LlambayMasset2016} which is the successor of the hydrodynamic FARGO code. Both codes use the orbital advection algorithm of \citet{Masset2000}, which 
significantly increases the timestep in thin protoplanetary discs. 

We use a reference frame centred on the star,  and corotating with the planet. 
The gravitational potential $\Phi$ due to the central star and the planet is given by
\begin{equation}
       \Phi=\Phi_S+\Phi_p,
	\label{eq:potential}
\end{equation}
where
\begin{equation}
       \Phi_S=-\frac{GM_S}{r},
	\label{eq:star_p}
\end{equation}
and
\begin{equation}
       \Phi_p=-\frac{GM_p}{\sqrt{r_p^2+\epsilon^2}}+\frac{GM_pr\cos\phi}{r_p^2},
	\label{eq:planet_p}
\end{equation}
where $r_{p}\equiv |\vecr-\vecR_{p}|$ is the distance  from the
planet, and $\epsilon$ is a softening 
length used to avoid computational problems arising from a divergence of the potential in the vicinity of the planet. We use $\epsilon=0.6H_{p}$, where $H_{p}$ is the disc scaleheight at $R=R_{p}$, but
we also performed simulations with $\epsilon=0.3 H_{p}$
to check that the results are not sensitive to the exact value of $\epsilon$.
The last term in Equation (\ref{eq:planet_p}) is the indirect term which appears 
because the reference frame is non-inertial and centred on the star.
The self-gravity of the disc is ignored in our simulations.

In Runs 1 to 10, the planet, whose orbit is tilted by an angle $i$ with respect to the initial midplane
of the disc, is forced to describe a circular orbit of radius $R_{p}$
around the central star. Thus, we ignore the changes in the orbital parameters
of the planet caused by tidal torques (see \S\ref{sec:freeplanet} for simulations of a planet 
that is left to freely migrate because of the tidal torques).
No accretion of gas by the planet is considered here. 
 
The aspect ratio, $h\equiv H/r$, is assumed to be constant across the disc, where $H$ is the vertical 
scale height of the disc. The initial density of the disc, $\rho_{i}(R,z)$, is derived by assuming
a power-law surface density
\begin{equation}
\Sigma_{i}(R)=\Sigma_{0}\left(\frac{R_{p}}{R}\right)^{1/2},
\end{equation}
and by imposing hydrostatic equilibrium, which
implies that $c_s(r)=hr\Omega$ where $\Omega(r)$ is the Keplerian angular velocity around the star.
Note that $\Sigma_{0}$ denotes the initial surface density at $R=R_{p}$.

Our distance unit is $R_p$ and our time unit $\omega^{-1}$ (as defined
in section~\ref{sec:cop_case}, $\omega$ is the angular
velocity of the planet). The period of the planet is therefore $2\pi$.

The domain of the simulations extends radially from $r=0.4$ to $r=2.5$. The polar angle $\theta$ 
covers $14^{\circ}$ (from $83^{\circ}$ to $97^{\circ}$) for the simulations
with $h=0.05$ and $8^{\circ}$ (from $86^{\circ}$ to $94^{\circ}$)
for discs with $h=0.025$. 
All the simulations have the same grid size $(N_{r},N_{\phi},N_{\theta})=(266, 768, 64)$.

In the radial direction, we implemented damping boundary conditions for the
radial component of the velocity $v_{r}$ \citep{Deval2006}.
More specifically, $v_{r}$ is artificially damped in the regions $r\in [0.4, 0.5]$
and $r\in [2.1, 2.5]$ by solving the equation
\begin{equation}
       \frac{\partial v_r}{\partial t}=-\frac{\Omega}{2\pi}\left(v_r-v_{r0}\right)\chi(r)
	\label{eq:Stockholm}
\end{equation}
after each time-step. Here $v_{r0}$ is the radial velocity component at $t=0$ and $\chi(r)$ is a 
parabolic function which takes the value $1$ at the domain boundary and $0$ at
the limit of the damping region \citep[e.g.,][]{Deval2006}. For the other two velocity 
components and the density we use reflecting boundary conditions without any damping.

\subsection{Results}
In this Section, we investigate numerically the interaction of an inclined 
planet with the disc. 
The parameters of the simulations for planets
on fixed orbits are given in Table \ref{tab:simulations}. In \S \ref{sec:freeplanet} we present
a simulation  for a freely moving planet. In most of our simulations, we
take $q=10^{-3}$, $h=0.05$ and a kinematic viscosity $\nu=10^{-5}$.

\subsubsection{Evolution of the disc: planets on fixed orbits}
Here we study the formation and structure of the gap carved by 
a planet on a fixed inclined circular orbit for several inclinations
$0^{\circ},4^{\circ}, 10^{\circ}, 15^{\circ}, 20^{\circ}$, and $30^{\circ}$
with respect to the initial midplane of the disc, which corresponds to the plane $\theta = 90^\circ$.
The simulations were run over $200$ planetary orbits.
The strengh of the interaction between the planet and the disc depends
on the inclination of the planetary orbit.
For $h=0.05$, the aperture angle of the disc is $2.9^{\circ}$. For
inclinations larger than $2.9^{\circ}$, 
the planet spends a fraction of its orbital period within the disc
that decreases as the inclination increases.
Figures \ref{fig:disc_integer} and \ref{fig:disc1} display the
perturbation of volume
density after $200$ orbits  on equatorial and meridional cross sections, for 
$q=10^{-3}$, $h=0.05$, $\nu=10^{-5}$ and different inclinations.
As occurs in the coplanar case, the planet triggers a wake with two spiral arms that emanate
from the planet (see Figures \ref{fig:disc_integer}$a$ and \ref{fig:disc1}$a$).  A gap in the surface density is also apparent.
The gap divides the disc into two regions: $i)$ the inner disc, $r<R_p$, and the outer disc at $r>R_{p}$.
In the inner (outer) disc, the spiral arm is leading (trailing), as in the non-inclined case.
The amplitude of the spiral arms depends on the inclination of the planetary orbit as can be observed in the cases $a)-f)$ in Figure \ref{fig:disc_integer}.
Figure \ref{fig:wake} shows the perturbed density along the crest of the outer
spiral arm. It can be seen that for $i=4^{\circ}$ the density enhancement along the crest
is three times greater than for $i=20^{\circ}$. 

Since the timescale for gap opening is much larger than the dynamical
timescale, the structure of the gap is rather axisymmetric (i.e. it does
not depend on the phase of the planet), except in the vicinity of planet's
position. 
A magnification of the density map in a box
centred at $x=X_{p}$, $y=Y_{p}$ and $z=0$ is shown in Figure \ref{fig:disk_zoom}.
Note that the planet is at its maximum distance from the disc in these
snapshots. We see that the higher inclination case has a less clean gap,
and finer substructure. For $i=20^{\circ}$, the spiral waves do not
emanate from the projected position of the planet. 

The appearance of the gap of an inclined planet is somehow reminiscent
of the aspect of HL~Tau gap structures \citep[][]{ALMAPartnershipetal2015, ALMAPartnershipetal2015a, ALMAPartnershipetal2015b, ALMAPartnershipetal2015c}, in
which we do not see local, conspicuous density enhancements at
particular azimuth, which could potentially be due to planets. A large
number of mechanisms can account for the existence of the gap
structures observed in HL~Tau \citep[][]{Carrasco-Gonzalezetal2009, Flocketal2015, Gonzalezetal2015, Zhangetal2015, Carrasco-Gonzalezetal2016, Okuzumietal2016, Rugeetal2016, Hsi-WeiYenetal2016}. 
We simply mention
that the lack of localized structures within the gap does not rule out
planetary torques as the mechanism responsible for their existence,
since a midly inclined planet does not trigger a large density
enhancement at its projected location in the gap.

The tidal perturbation of a planet with non-zero inclination may lead to the excitation of
vertical disturbances (bending waves or warps) in the disc  
\citep[for a binary star, see, e.g.,][]{PapaloizouTerquem1995, Larwoodetal1996}.
For massive enough planets, the disc will try to realign with the orbital plane of the planet, reducing
the relative inclination between planet's orbit and the disc.
In order to quantify the excitation of vertical modes (warps) in the disc,
we calculated the inclination of the disc at different radii using the expression
\begin{equation}
       i_{\textsc{\tiny D}}(r)=\arccos{\frac{L_z(r)}{\abs{\vecL(r)}}},
	\label{eq:discinc}
\end{equation}
where $\vecL (r)$ is the angular momentum vector of a differential ring of the disc:
\begin{equation}
\vecL (r)= \int \rho (\vecr \times \vecv) \,d\theta \,d\phi.
\end{equation}
For the adopted values of $q$, the disc hardly changes its inclination (Figure \ref{fig:inclination}). 
The largest values of $i_{\textsc{\tiny D}}$ occur for the case $i=4^{\circ}$ and for rings
having radii $\simeq R_{p}$. In particular, for $q=10^{-3}$, $h=0.05$ and 
$i=4^{\circ}$, these rings are displaced by an angle $\simeq 2^{\circ}$. 
For $i\geq 10^{\circ}$, $i_{\textsc{\tiny D}}$ is very small compared to $i$ and hence it is a good 
approximation to ignore the bend of the disc.

\subsubsection{Scaling relations for the gap}
\label{sec:scaling}
As expected, the depth of the gap depends on the planetary inclination $i$. 
At low inclinations ($i<10^{\circ}$), the time for gap opening is shorter than in the case 
where the planet orbit is more inclined.
As a measure of the depth, we determine $\Sigma_{\rm gap}$ from our
simulations by calculating the surface density averaged over 
azimuth and over the radial direction between $R_p-\sqrt{2}R_H$ and $R_p+\sqrt{2}R_H$.
Figure \ref{fig:gapsito} displays $\Sigma_{\rm gap}$ as a function of time for Runs 1 to 6.
The gap density $\Sigma_{\rm gap}(t)$ converges toward a constant value at larger time. 
In order to compare the rate of emptying of gas in the gap,
we write 
\begin{equation}
\Sigma_{\rm gap}(t)=\left[1-f(t)\right]\Sigma_{0}+f(t) \Sigma_{200},
\end{equation}
where $\Sigma_{200}$ denotes the surface density in the gap at $t=200$ orbits
and $f(t)$ is an auxiliary function that satisfies $f(0)=0$ and $f(200)=1$.
If $f(t)$ becomes flat, it means that we have essentially reached the asymptotic
value of $\Sigma_{\rm gap}$. 
As judged from Figure \ref{fig:gapsito}, the rate of depletion of gas in the gap is very
low after $200$ orbits, indicating that $\Sigma_{200}$ may be considered
as representative of the asymptotic value, except perhaps for the simulation
with $i=30^{\circ}$. In fact, the gap cleaning proceeds slightly slower for high $i$
(lower panel in Figure \ref{fig:gapsito}).

In order to compare how the process of gap opening occurs for different 
planetary inclinations, we plot the 
 azimuthally-averaged surface density of the gap at the time when 
the condition $\Sigma_{\rm gap}=\Sigma_0/2$ is satisfied (Figure \ref{fig:S2}).
For $i=4^{\circ}, 10^{\circ}, 15^{\circ}$, and $20^{\circ}$ this occurs
at $25$, $40$, $50$, and $109$ orbits, respectively. 
In the case $i=30^{\circ}$,  the planet is unable to carve such a deep gap during the time of the 
simulation ($200$ orbits); at the end of this simulation
$\Sigma_{\rm gap}=0.67\Sigma_{0}$. We see that at the time when 
$\Sigma_{\rm gap}=\Sigma_{0}/2$, the profile of the gaps are clearly
different. For $i=4^{\circ}$, the planet satisfies the condition
$\Sigma_{\rm gap}=\Sigma_{0}/2$ in a shorter timescale and, moreover,
it depletes more material, leading to a wider gap, than planets with larger
inclinations. This means that the gap cleaning is more efficient, along 
a wider radial range, for low inclination planets. 
Also, the local maxima of the density that appear near the edges of the
evacuated gap have more time to spread by viscous diffusion, since the
plots of larger $i$ are made at a later time.

For each simulation, Table \ref{tab:simulations} lists the values of $q_{\rm crit}$, as derived in 
\S \ref{sec:criterion_i}. We see that the
viscous criterion for gap formation is roughly satisfied, 
in the sense that when $q>q_{\rm crit}$ it holds that $\Sigma_{200}\lesssim 0.2\Sigma_{0}$.

Figure \ref{fig:sigma1} shows the radial profile of the azimuthally-averaged surface density after 
$45$ and $200$ planetary orbits and different inclination angles (but having the same kinematic
viscosity and aspect ratio). It is apparent that the depth
of the gap decreases with the inclination angle of the planet's orbit.
The surface density bumps at $r=1.4$ at $t=45$ and $t=200$ orbits appear because the profiles of the
surface density are not completely relaxed at $t\leq 200$ orbits given that 
the viscous timescale is $\sim 2.5\times 10^{3}$ orbits.

We have compared the resultant gap profiles in the simulations
with those predicted using the 1D model (the method
described in \S \ref{sec:timevolution}), which assumes that the wake's torque is deposited
locally in the disc. 
We use $l_{\rm min}=(q/3)^{1/3}$ (see \S \ref{sec:cutoffs}) and explore two values 
for $\xi$ ($\xi=1$ and $\xi=2$).
We also plot the steady-state gap profile by integrating
numerically Equation (\ref{eq:Sigma_diff}), with the boundary condition $\Sigma (2.5)=\Sigma_{0}/\sqrt{2.5}$, 
which is the unperturbed surface density at $R=2.5R_p$, and using $r_{\rm min}=R_{H}$.

To be fully satisfactory, the models should be able to reproduce the
width and depth of the gap at any time.
From Figure \ref{fig:sigma1}, we see that the width of the gaps is fairly reproduced
for both $\xi=1$ and $\xi=2$. However, models with $\xi=1$ predict
shallower gaps than those found in the simulations for inclinations
$\geq 10^{\circ}$. 
It is remarkable that the depth of the gaps at $200$ orbits in the full 3D simulations 
is larger than the depth of the stationary gaps in the 1D model,
i.e. the `steady state' curves in Figure \ref{fig:sigma1}. This suggests that
the value of $\xi$ is larger than $1$. This is not unexpected if one
reminds that the torque calculated by summing the contribution from all
Lindblad resonances in the coplanar case is a factor of $2.8$ larger
than the torque calculated with the impulse approximation [see \S \ref{sec:cop_case} and 
\citet{Lin1986a}].

Adopting $\xi=2$, there is a good level of agreement between the gap profiles derived using 
the 1D model and those found in the simulations 
for inclinations between $10^{\circ}$ and $20^{\circ}$ (see Figure \ref{fig:sigma1}). 
For lower inclinations ($i<10^{\circ}$), the 1D model overestimates the depth of the gap 
when assuming our fiducial value of $l_{\rm min}$ (not shown).
In order to reproduce the gap profile found in the simulation with $i=4^{\circ}$, 
we need $l_{\rm min}=1.8(q/3)^{1/3}$.

For $i=30^{\circ}$, the 1D model clearly underestimates the depth of the gap
at any time (Figure \ref{fig:sigma1}), indicating that at least one of our assumptions is not
fully correct. It is plausible that for inclinations as large as $30^{\circ}$,
the impulse approximation underestimates the torque. Moreover, the local
damping approximation is less justified as the inclination increases
because the perturbed density in the wake decreases (Figure \ref{fig:wake}). However,
it is unclear that the resulting angular momentum flux driven by
spiral arms will increase gap clearing in the vicinity of the planet.

In order to explore a bit further the 1D model, 
Figure \ref{fig:diffm} compares the gap profiles in our
full 3D simulations with those in 1D models
for $i=20^{\circ}$ and two different values of $q$. We see that for $q=3\times 10^{-3}$,
the 1D model is still successful in reproducing the gap profile. However,
for $q=3\times 10^{-4}$, the evacuation of gas from the gap is more efficient
in the full 3D simulations than in the 1D models.

\begin{figure}
	\includegraphics[width=\columnwidth]{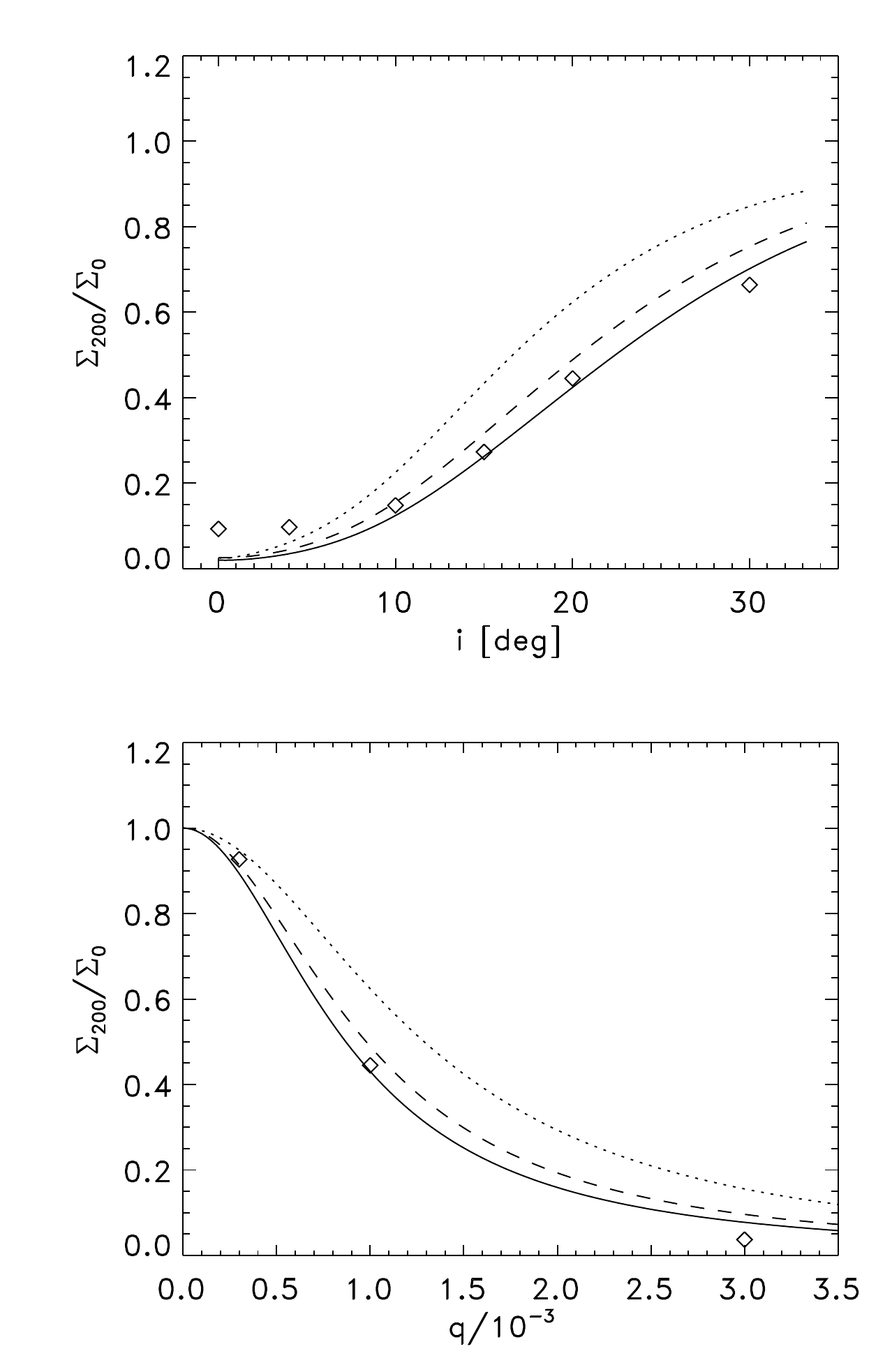}
    \caption{Predicted values of $\Sigma_{200}/\Sigma_{0}$ using
the zero-dimensional approach (see \S \ref{sec:method2}) with $l_{\rm min}=0.87h$ 
(dotted lines), $l_{\rm min}=1.1h$ (dashed lines) and $l_{\rm min}=1.2h$ (solid lines),
together with the values from
the numerical simulations (symbols) for simulations with different inclinations
(Run 1 to Run 6; top panel) and for simulations with different $q$
(Runs 5, 8 and 9; bottom panel). }
 \label{fig:duffell_i}
\end{figure}

It is worthwhile to consider the predictions using the `zero-dimensional' approximation.
Following the procedure described in Section \ref{sec:method2}, we have calculated
$\hat{f}_{0}(i,h)$ for $l_{\rm min}=0.87h, 1.1h$ and $1.2h$. 
Figure \ref{fig:duffell_i} compares $\Sigma_{\rm gap}$ calculated using 
Equation (\ref{eq:duffell_eq}) with the values obtained in the simulations. 
The general trend that the gap is shallower when increasing $i$ is consistent with simulations. 
A value for $l_{\rm min}$ of $1.2h$ is required to reproduce the gap density values for a disc
with aspect ratio $=0.05$.

It is worthwile to look at Run 10. This simulation has $h=0.025$ and lies in the deeply nonlinear regime.
For this run, the zero-dimensional approximation with $l_{\rm min}=1.2h$ predicts 
$\Sigma_{200}/\Sigma_{0}=0.38$,
while the simulated disc has a gap with $\Sigma_{200}/\Sigma_{0}=0.26$. 
This illustrates that hydrodynamical effects may be important for very thin discs.

\subsection{Radial migration and inclination damping for free planets}
\label{sec:freeplanet}

In the impulse approximation, it is possible to estimate the characteristic
timescales for radial migration and inclination damping for planets with
inclination large enough that they cross the disc at supersonic velocities.
\citet{Rein2012} find that the inclination and the semimajor axis damping
timescales, measured in units of the orbital period, are
\begin{equation}
\tau_{\rm inc}\equiv i/(2|di/dt|)= \frac{M_{S}} {2\pi q\Sigma_{0}R_{p}^{2}}
\frac{i\sin^{3}(i/2)}{\ln\Lambda},
\label{eq:tau_inc}
\end{equation}
\begin{equation}
\tau_{\rm R}\equiv R_{p}/(2|\dot{R}_{p}|)=\frac{M_{S}}{8\pi q\Sigma_{0}R_{p}^{2}}
\frac{\sin(i/2)\sin(i)}{\ln\Lambda},
\label{eq:tau_R}
\end{equation}
where  
$\Sigma_{0}$ is the surface density of the disc at the intersection of the planetary orbit with the disc,
and $\ln\Lambda$ is the Coulomb logarithm of the interaction. More specifically, $\Lambda$ is
the ratio between the upper ($r_{\rm max}$) and lower ($r_{\rm min})$ cut-off length scales of the interaction.
A similar formula (except by a factor of $2$) for the timescale for the orbit to change was derived 
by \citet{Xiang2013}. 
The timescales $\tau_{\rm inc}$ and $\tau_{\rm R}$ depend on the unperturbed
local surface density of the disc, $\Sigma_{0}$, because it was assumed that the timescale
for gap opening is larger than both $\tau_{\rm inc}$ and $\tau_{\rm R}$.
If the planet is able to open a gap, the 
depletion of material in the planet vicinity
should be taken into account. Once the planet has carved a gap, the rates of damping
in $i$ and $R$ are expected to decrease \citep[e.g.,][]{Xiang2013}.

\begin{figure}
	\includegraphics[width=\columnwidth]{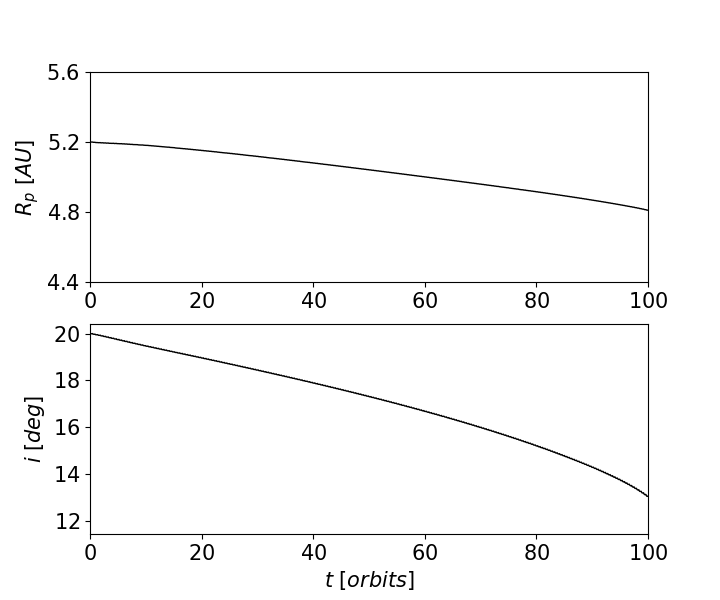}
    \caption{Top: Semimajor axis $R_p$ as function of time for a planet with $q=10^{-3}$
and an initial inclination of $i_0=20^{\circ}$. The disc has $\Sigma_{0}=210$ g cm$^{-2}$. 
Bottom: Temporal evolution of the planetary inclination $i$.}
 \label{fig:damp}
\end{figure}

The timescale for inclination damping may be comparable or even
smaller than the timescale for gap opening if the disc is sufficiently massive. 
For illustration, consider a planet-star system with $q=10^{-3}$, $M_{S}=1M_{\odot}$ and
$i=20^{\circ}$. For these parameters, $\tau_{\rm inc}$ is given by
\begin{equation}
\tau_{\rm inc}=\frac{450}{\ln\Lambda} \left(\frac{\Sigma_{0}}{200\,{\rm g \,cm}^{-2}}\right)^{-1}
\left(\frac{R_{p}}{5.2\,{\rm AU}}\right)^{-2} \,\,{\rm orbits}.
\label{eq:tau_inc20}
\end{equation}
Since the timescale to form a gap with a depth $\Sigma_{\rm gap}=0.5\Sigma_{0}$ is $128$ orbits
(see Figure \ref{fig:S2}),
the inclination damping timescale is comparable to or smaller than the gap opening timescale if
$\Sigma_{0}\gtrsim 600/\ln\Lambda$ g cm$^{-2}$ (assuming $R_{p}=5.2$ AU, $i=20^{\circ}$
and $q=10^{-3}$). The critical surface density is expected to increase linearly with $q$ because 
$\tau_{\rm inc}\propto q^{-1}$, whereas the timescale for gap opening goes as 
$\sim q^{-2}$.

In order to test these estimates, we perform one simulation in which the planet feels the tidal 
torques by the disc since the beginning of the simulation. 
The planet is initially in a circular orbit with $R_{p}=5.2$ AU and $i_{0}=20^{\circ}$.
It has $q=10^{-3}$ and a softening radius $\epsilon=0.3H_{p}$.
The initial surface density at $5.2$ AU is $210$ g cm$^{-2}$. Figure \ref{fig:damp}
shows the temporal evolution of $R_{p}$ and $i$. In $100$ orbits, the semimajor axis decays
from $5.2$ AU to $4.8$ AU and the inclination from $20$ to
$13$ deg, resulting in $di/dt=-0.07$ deg/orbit and $dR_{p}/dt=-4\times 10^{-3}$ AU/orbit.
These rates of inclination damping and radial migration are consistent with those found in previous studies 
\citep[][]{MarzariNelson2009,Xiang2013}.
For instance, for a planet with $i_{0}=20^{\circ}$, $\Sigma_{0}=76$, and $q=10^{-3}$,
\citet{Xiang2013} find that $di/dt=-0.028$ deg/orbit and $dR_{p}/dt=-0.9\times 10^{-3}$ AU/orbit.

In order to compare with the predictions of Equations (\ref{eq:tau_inc})-(\ref{eq:tau_R}),
we need to estimate $\ln\Lambda$ in our simulation. To do so, we use that 
$r_{\rm min}$ depends on the softening radius $\epsilon$ as
$r_{\rm min}\simeq 2.25\epsilon = 0.67H_{p}$ 
\citep{Bernal2013} and that $r_{\rm max}\simeq 2.1\sqrt{2} H_{p}$ in a disc \citep{Canto2013,Xiang2013}.
Hence, we find that $\ln \Lambda\simeq 1.5$. Using this value, Equations (\ref{eq:tau_inc}) and (\ref{eq:tau_R})
predict $di/dt=-0.027$ deg/orbit and $dR_{p}/dt=-0.9\times 10^{-3}$ AU/orbit.
These values are a factor of $3-4$ smaller than those found in the simulations.
Nevertheless, it is likely that the accuracy is better for larger inclinations \citep{Rein2012,Xiang2013}.

\begin{figure}
	\includegraphics[width=\columnwidth]{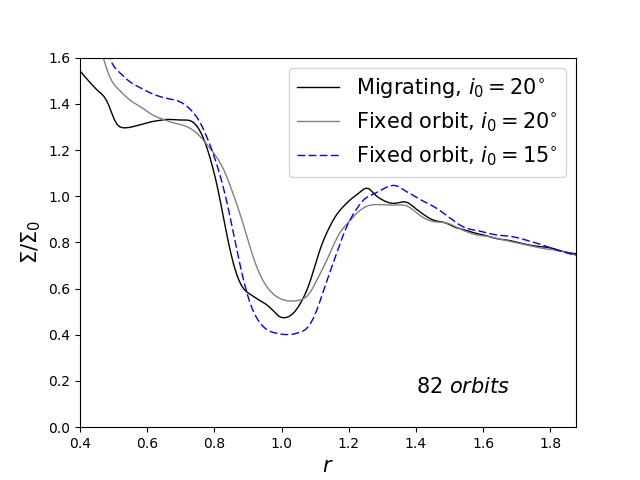}
    \caption{Comparison of the radial profiles of the surface density for migrating and non-migrating
planets, after $82$ orbits.}
 \label{fig:freeSigma}
\end{figure}

Given that the inclination damping timescale is comparable to the timescale for gap clearing,
the process is dynamical in the sense that the inclination cannot be assumed to be constant.
Figure \ref{fig:freeSigma} plots the surface density profile after $82$ orbits, that is, when the planet has
an inclination of $15^{\circ}$, together with the surface density profile in simulations where
the planets are at fixed orbits with inclinations $15^{\circ}$ and $20^{\circ}$. As expected, the surface
density has its minimum at a inner radius when the planet is allowed to migrate.
In addition, for the planet starting with $i_{0}=20^{\circ}$,
the depth of the gap is larger when the planet is free to migrate than when the planet
is forced to orbit at constant inclination ($20^{\circ}$). However, the planet that is forced 
to orbit at a constant inclination of $15^{\circ}$ opens a deeper gap after $82$ orbits than
the gap produced by the migrating planet. 

We conclude that for planets with masses of the order of $M_{J}$ and for 
$\Sigma_{0}\gtrsim 100$ g cm$^{-2}$, the inclination damping timescale is comparable to or shorter than
the gap clearing process and, therefore, it is necessary to solve the time-dependent 1D model
described in \S\ref{sec:timevolution}. For those values of $\Sigma_{0}$, the inclination is damped 
to zero in a timescale much shorter than the lifetime of the disc. Therefore, the scattering process of inclined
planets should occur when the surface density of the disc was significantly smaller than
$100$ g cm$^{-2}$ \citep[see also][]{Bitsch2013}.

\section{Conclusions}
 \label{sec:conclusions}

We have developed a model to understand the dynamical response of a protoplanetary disc
to the presence of a planet in inclined orbit. We considered planets massive enough
to open a gap but not too massive to warp the disc significantly. 
Given that the impulse approximation for non-inclined planets yields
correct scalings and better than a factor of $2$ estimates of the
torque \citep[e.g.][]{Lin1993,Armitage2010}, we have computed the excitation torque density by
inclined planets on circular orbits in the impulse approximation. 
Using this simple approach, we have derived a viscous criterion for the 
formation of gaps by mildly inclined planets ($i\leq 30^{\circ}$) [see \S \ref{sec:criterion_i}].
Such a criterion may be useful when the planet has acquired
its inclination after the gas in the disc is well depleted, or to interpret simulations that are started after
the disc has evolved with the planet at a fixed, constant inclination.

For planets that are forced to describe fixed circular orbits, we have calculated 
the temporal evolution of the gap profile in the impulse approximation and 
using the hypothesis of local damping. We have compared these radial gap profiles with those
derived in 3D hydrodynamical simulations. The simple model underestimates the depth
of the gap for $i\geq 10^{\circ}$ when comparing with the results of our
simulations. Introducing a correction factor of $2$ in the torque allows us
to reproduce successfully the temporal evolution of the gap profile for
inclinations between $10^{\circ}$ and $20^{\circ}$, and planetary 
masses $\geq 1M_{J}$. For planetary inclinations
larger than $20^{\circ}$, the simple model underestimates the depth of the gap
probably because one assumption on which the impulse approximation is based,
namely that the interaction is local, is only poorly verified at large inclinations.

We have also computed the depth of the stationary gap in the so-called
zero-dimensional approximation and find that it accounts correctly the trend
of the gap depth with the inclination and mass of the planet.

In order to check the validity of the approximations made in our approach, we have mainly 
focused on planets in fixed circular orbits. This approximation is strictly valid only if the 
inclination damping timescale is larger than the timescale for gap opening. Nevertheless, 
for given, arbitary functions $i(t)$, our formalism allows to derive the gap profile as a function
of  time. The results will be most accurate for $10^{\circ}\leq i(t) \leq 20^{\circ}$.

 \section*{Acknowledgements}
We thank the referee for useful comments which improved the paper appreciably.
The computer Tycho 2 (Posgrado en Astrof\'{\i}sica-UNAM, Instituto de Astronom\'{\i}a-UNAM
and PNPC-CONACyT) has been used in this research.
This work has been partially supported by CONACyT grant 165584, SIP
20161416 and UNAM's DGAPA grant PAPIIT IN101616.




\bibliographystyle{mnras}




\begin{thebibliography}{99}

\bibitem[\protect\citeauthoryear{Albrecht et al.}{2012}]{Albrecht2012} Albrecht S. et al., 2012, \apj, 757, 18

\bibitem[\protect\citeauthoryear{ALMA Partnership et al.}{2015}]{ALMAPartnershipetal2015} ALMA Partnership, Fomalont, E. B., Vlahakis C., et al., 2015, \apj, 808, L1

\bibitem[\protect\citeauthoryear{ALMA Partnership et al.}{2015a}]{ALMAPartnershipetal2015a} ALMA Partnership, Hunter, T. R., Kneissl R., et al., 2015a, \apj, 808, L2

\bibitem[\protect\citeauthoryear{ALMA Partnership et al.}{2015b}]{ALMAPartnershipetal2015b} ALMA Partnership, Brogan, C. L., Perez L. M., et al., 2015b, \apj, 808, L3

\bibitem[\protect\citeauthoryear{ALMA Partnership et al.}{2015c}]{ALMAPartnershipetal2015c} ALMA Partnership, Vlahakis C., Hunter T. R., et al., 2015c, \apj, 808, L4


\bibitem[\protect\citeauthoryear{Armitage}{2010}]{Armitage2010}Armitage P. J. 2010, 
Astrophysics of planet formation, Cambridge University Press

\bibitem[\protect\citeauthoryear{Baruteau \& Masset}{2013}]{Baruteau2013}
Baruteau C., Masset F.  2013, Lecture Notes in Physics, Vol. 861, Tides in Astronomy and
Astrophysics, Springer-Verlag, Berlin, p. 201

\bibitem[\protect\citeauthoryear{Ben\'{\i}tez-Llambay \& Masset }{2016}]{Benitez-LlambayMasset2016} Ben\'{\i}tez-Llambay P., Masset F. S. 2016, ApJS, 223, 11

\bibitem[\protect\citeauthoryear{Bernal \& S\'anchez-Salcedo}{2013}]{Bernal2013}
Bernal C. G., S\'anchez-Salcedo F. J. 2013, \apj, 775, 72

\bibitem[\protect\citeauthoryear{Bitsch \& Kley}{2011}]{BitschKley2011}Bitsch B., Kley W. 2011,
\aap, 530, 41 

\bibitem[\protect\citeauthoryear{Bitsch et al.}{2013}]{Bitsch2013}Bitsch B., Crida A.,
Libert A.-S., Lega E. 2013, \aap, 555, 124 

\bibitem[\protect\citeauthoryear{Bryden et al.}{ 1999}]{Brydenetal1999}Bryden G., Chen X., Lin D. N. C., Nelson R. P., Papaloizou J. C. B. 1999, ApJ, 514, 344
\bibitem[\protect\citeauthoryear{Cant\'o et al.}{2013}]{Canto2013}
Cant\'o J., Esquivel A., S\'anchez-Salcedo F. J., Raga A. C. 2013, \mnras, 762, 21

\bibitem[\protect\citeauthoryear{Carrasco-Gonzalez et al.}{ 2009}]{Carrasco-Gonzalezetal2009}Carrasco-Gonzalez C., Rodriguez L. F., Anglada G., Curiel S.  2009, \apj, 693, L86

\bibitem[\protect\citeauthoryear{Carrasco-Gonzalez et al.}{ 2016}]{Carrasco-Gonzalezetal2016}Carrasco-Gonzalez C., et al. 2016, ApJ, 821, L16
\bibitem[\protect\citeauthoryear{Cresswell \& Nelson }{2006}]{CresswellNelson2006}Cresswell P., Nelson R. P. 2006, A\&A, 450, 833

\bibitem[\protect\citeauthoryear{ Cresswell et al.}{ 2007}]{Cresswelletal2007}Cresswell P., Dirksen G., Kley W.,  Nelson R. P. 2007, A\&A, 473, 329
\bibitem[\protect\citeauthoryear{Crida et al.}{ 2006}]{Cridaetal2006}Crida A., Morbidelli A., Masset F. 2006, Icarus, 181, 587

\bibitem[\protect\citeauthoryear{de Val-Borro et al.}{ 2006}]{Deval2006}de Val-Borro M. et al. 2006, 
MNRAS, 370, 529

\bibitem[\protect\citeauthoryear{Duffell et al.}{ 2014}]{Duffelletal2014}Duffell P. C., Haiman Z.,
MacFadyen A. I., D'Orazio D. J., Farris B. D. 2014, ApJ, 792, L10

\bibitem[\protect\citeauthoryear{Duffell}{2015}]{Duffell2015}
Duffell P. C. 2015, \apj, 807, L11

\bibitem[\protect\citeauthoryear{Fabrycky \& Winn}{2009}]{Fabrycky2009}Fabrycky D. C.,  Winn J. N. 2009, ApJ, 696, 1230

\bibitem[\protect\citeauthoryear{Flock et al.}{ 2015}]{Flocketal2015}Flock M., Ruge J. P., Dzyurkevich N., Henning Th., Klahr H., Wolf S. 2015, A\&A, 574, A68

\bibitem[\protect\citeauthoryear{Fung et al.}{ 2014}]{Fungetal2014}Fung J., Shi J.-J., Chiang E. 2014, ApJ, 782, 88

\bibitem[\protect\citeauthoryear{Goldreich \& Tremaine}{1980}]{GoldreichTremaine1980}Goldreich P.,  Tremaine S. 1980, ApJ, 241, 425

\bibitem[\protect\citeauthoryear{Gonzalez et al.}{2015}]{Gonzalezetal2015}
Gonzalez J. F., Laibe G., Maddison S. T., Pinte C., M\'enard F. 2015, \mnras, 454, L36

\bibitem[\protect\citeauthoryear{Hosseinbor et al.}{2007}]{hos07} Hosseinbor A. P., Edgar R. G.,
Quillen A. C., LaPage A. 2007, \mnras, 378, 966

\bibitem[\protect\citeauthoryear{Kanagawa et al.}{2015}]{Kanagawaetal2015}
Kanagawa K. D., Tanaka H., Muto T., Tanigawa T., Takeuchi T. 2015, \mnras, 448, 994

\bibitem[\protect\citeauthoryear{Kley et al.}{ 2000}]{Kleyetal2000}Kley W., D'Angelo G., Henning T. 2000, ApJ, 547, 447
\bibitem[\protect\citeauthoryear{Kley \& Nelson}{2012}]{Kley2012}
Kley W., Nelson R. P. 2012, \araa, 50, 211

\bibitem[\protect\citeauthoryear{Larwood et al.}{ 1996}]{Larwoodetal1996}
Larwood J. D., Nelson R. P., Papaloizou J. C. B., Terquem C. 1996, MNRAS, 282, 597

\bibitem[\protect\citeauthoryear{Lin \& Papaloizou }{ 1979}]{LinPapaloizou1979}Lin D. N. C.,  Papaloizou J. C. B. 1979, MNRAS, 186, 799 
\bibitem[\protect\citeauthoryear{Lin \& Papaloizou}{1986a}]{Lin1986a}
Lin D. N. C., Papaloizou J. C. B. 1986a, \apj, 307, 395
\bibitem[\protect\citeauthoryear{Lin \& Papaloizou}{1986b}]{Lin1986b}
Lin D. N. C., Papaloizou J. C. B. 1986b, \apj, 309, 846
\bibitem[\protect\citeauthoryear{Lin \& Papaloizou}{1993}]{Lin1993}
Lin D. N. C.,  Papaloizou J. C. B. 1993, in Protostars and Planets III, ed. E. H. Levy \& J. I. Lunine (Tuczon, AZ: Univ. Arizona Press), 749-835 
\bibitem[\protect\citeauthoryear{Lubow et al.}{2015}]{Lubow2015}
Lubow S. H., Martin R. G., Nixon C. 2015, \apj, 800, 96
\bibitem[\protect\citeauthoryear{Malik et al.}{2015}]{Malik2015}
Malik M., Meru F., Mayer L., Meyer M. 2015, \apj, 802, 56

\bibitem[\protect\citeauthoryear{Marzari \& Nelson}{2009}]{MarzariNelson2009}Marzari F.,  Nelson A. F. 2009, ApJ, 705, 1575

\bibitem[\protect\citeauthoryear{Masset}{2000}]{Masset2000}Masset F. S. 2000, A\&AS, 141, 165 

\bibitem[\protect\citeauthoryear{Mayer et al.}{2002}]{Mayeretal2002}Mayer L.,  Quinn T., Wadsley J., Stadel J., 2002, Science, 298, 1756 
\bibitem[\protect\citeauthoryear{Miranda \& Lai}{2015}]{Miranda2015}
Miranda R., Lai D. 2015, \mnras, 452, 2396
 

\bibitem[\protect\citeauthoryear{Okuzumi et al.}{2016}]{Okuzumietal2016}
Okuzumi S., Momose M., Sirono S., Kobayashi H., Tanaka H. 2016, arXiv:1510.03556

\bibitem[\protect\citeauthoryear{Papaloizou \& Lin}{1984}]{Papaloizou1984}
Papaloizou J., Lin D. N. C. 1984, \apj, 285, 818

\bibitem[\protect\citeauthoryear{Papaloizou \& Terquem}{1995}]{PapaloizouTerquem1995}
Papaloizou J. C. B.,  Terquem C. 1995, MNRAS, 274, 987

\bibitem[\protect\citeauthoryear{Picogna \& Marzari }{2015}]{Picogna2015}Picogna G., Marzari F. 2015, A\&A, 583, A133

\bibitem[\protect\citeauthoryear{Pollack et al.}{1996}]{Pollacketal1996}
Pollack J. B., Hubickyj O., Bodenheimer P., Lissauer J. J., Podolak M., Greenzweig Y., 1996, Icarus, 124, 62

\bibitem[\protect\citeauthoryear{Pringle}{1981}]{Pringle1981} Pringle J. E. 1981, \araa, 19, 137

\bibitem[\protect\citeauthoryear{Rein}{2012}]{Rein2012}Rein H. 2012, MNRAS, 422, 3611

\bibitem[\protect\citeauthoryear{Ruge et al.}{2016}]{Rugeetal2016}
Ruge J. P., Flock M., Wolf S., Dzyurkevich N., Fromang S., Henning Th., Klahr H., Meheut H., 2016,  A\&A, 590, A17 

\bibitem[\protect\citeauthoryear{Tanaka \& Ward }{2004}]{TanakaWard2004}Tanaka H., Ward W. R. 2004, ApJ, 602, 388
\bibitem[\protect\citeauthoryear{Thommes \& Lissauer}{2003}]{Thommes2003}
Thommes E. W., Lissauer J. J. 2003, \apj, 597, 566
\bibitem[\protect\citeauthoryear{Triaud et al.}{ 2010}]{Triaudetal2010}Triaud A. H. M., et al. 2010, A\&A, 524, A25
\bibitem[\protect\citeauthoryear{Varni\`ere et al.}{2004}]{Varniereetal2004}Varni\`ere P., Quillen A. C.,  Frank A. 2004, ApJ, 612, 1152
\bibitem[\protect\citeauthoryear{Ward}{1986}]{Ward1986} Ward W. R. 1986, Icarus, 67, 164
\bibitem[\protect\citeauthoryear{Ward \& Hourigan}{1989}]{Ward1989} Ward W. R., Hourigan K. 1989, \apj, 347,
490
\bibitem[\protect\citeauthoryear{Xiang-Gruess \& Papaloizou}{2013}]{Xiang2013}Xiang-Gruess M.,  
Papaloizou J. C. B. 2013, MNRAS, 431, 1320

\bibitem[\protect\citeauthoryear{Yen et al.}{ 2016}]{Hsi-WeiYenetal2016} Yen H.-S., Liu H. B.,
Gu P.-G., Hirano N., Lee C.-F., Puspitaningrum E., Takakuwa S.  2016, ApJ, 820, L25

\bibitem[\protect\citeauthoryear{Zhang et al.}{2015}]{Zhangetal2015}
Zhang K., Blake G. A., \& Bergin E. A. 2015, \apj, 806, L7










\end{thebibliography}



\appendix

\section{The prograde, coplanar orbit  as a particular case}

In the following, we specialize the equations given in \S \ref{sec:differential_ring} for
the case $i=0$ and $\varepsilon=1$.
Equations (\ref{eq:dist_min}) and (\ref{eq:vecv_rel}) reduce to:
\begin{equation}
\vecd_{\rm min}=\begin{pmatrix} [R_{d}-R_{p}]\cos \phi_{p} \\ [R_{d}-R_{p}]\sin \phi_{p} \\  0 \end{pmatrix},
\label{eq:dist_min0}
\end{equation}
and
\begin{equation}
\vecv_{\rm rel}=R_{p}\begin{pmatrix} [\omega -\Omega]\sin\phi_{p} \\ [\Omega -\omega] \cos \phi_{p} \\  0 \end{pmatrix},
\label{eq:vecv_rel0}
\end{equation}
respectively. Clearly, we have that $d_{\rm min}=|R_{d}-R_{p}|$ and $v_{\rm rel}=R_{p}|\omega-\Omega|$.

According to Equation (\ref{eq:delta_inc}) with $i=0$, the torque in an elementary ring is
\begin{equation}
\frac{dT_{g}(l)}{dR}=\frac{3\Sigma (R) \omega^{2} R_{p}^{3}|l|}{4\pi} 
 \int_{0}^{2\pi} ( {\mathcal{R}}_{1} \tilde{\vecv}_{\rm rel})\cdot \vece_{\phi} \,d\phi_{p}.
\label{eq:delta_inc0}
 \end{equation}
Remind that $|l|=|R_{d}-R_{p}|/R_{p}$, and ${\mathcal{R}}_{1}={\mathcal{R}}-{\mathcal{I}}$,
where ${\mathcal{R}}$ is the rotation matrix of angle $\delta_{e}$ around the direction of 
the vector $\vecd_{\rm min}\times \vecv_{\rm rel}$. For the coplanar case ($i=0$) and
for a Keplerian disc, the rotation  is around the vector $-\vece_{z}$, and 
the rotation matrix is given by
\begin{equation}
{\mathcal{R}}=\left( \begin{array}{ccc}
\cos \delta_{e} & \sin\delta_{e} & 0 \\
-\sin\delta_{e} & \cos\delta_{e} & 0 \\
0 & 0 & 1 \end{array} \right).
\end{equation}
For small values of $\delta_{e}$, we have
\begin{equation}
{\mathcal{R}}_{1}=\delta_{e}\left( \begin{array}{ccc}
-\delta_{e}/2 & 1 & 0 \\
-1 & -\delta_{e}/2 & 0 \\
0 & 0 & 0 \end{array} \right),
\end{equation}
and, consequently
\begin{equation}
( {\mathcal{R}}_{1} \tilde{\vecv}_{\rm rel})\cdot \vece_{\phi}=\frac{\omega-\Omega}{\omega}
\frac{\delta_{e}^{2}}{2}.
\label{eq:combination}
\end{equation}
On the other hand, the deflection angle is
\begin{equation}
\delta_{e}=\frac{4G^{2}M_{p}^{2}}{v_{\rm rel}^{4}d_{\rm min}^{2}}=
4q^{2}\left(\frac{\omega}{\omega-\Omega}\right)^{4}\left(\frac{R_{p}}{R_{d}-R_{p}}\right)^{2}.
\end{equation}
Substituting into Equation (\ref{eq:combination}) and using that 
$\omega-\Omega\simeq 3\omega (R_{d}-R_{p})/(2R_{p})$, we get
\begin{equation}
( {\mathcal{R}}_{1} \tilde{\vecv}_{\rm rel})\cdot \vece_{\phi}=\frac{16}{27}q^{2} 
\left(\frac{R_{p}}{R_{d}-R_{p}}\right)^{5},
\end{equation}
and substituing into Equation (\ref{eq:delta_inc0}) we finally obtain
\begin{equation}
\frac{dT_{g}}{dR}=\pm \frac{8}{9} q^{2}\Sigma R_{p}^{3}\omega^{2} \left(\frac{R_{p}}{R_{d}-R_{p}}\right)^{4},
\end{equation}
or 
\begin{equation}
\frac{dT_{g}}{dR}=\pm \frac{8}{9} q^{2}\Sigma R_{p}^{3}\omega^{2} \left(\frac{R_{p}}{\Delta}\right)^{4},
\end{equation}
where the positive sign is for the external disc ($R_{d}>R_{p}$) and the minus sign is for the
internal disc ($R_{d}<R_{p}$). Integrating the above equation over $\Delta$ between $\Delta_{0}$ and $\infty$,
we obtain
\begin{equation}
T_{g}=\pm \frac{8q^{2}\Sigma R_{p}^{4}\omega^{2}}{27}\left(\frac{R_{p}}{\Delta_{0}}\right)^{3}.
\end{equation}
We have recovered the same formula as given in \citet{Lin1993} (their equation 5).


\bsp	
\label{lastpage}
\end{document}